\documentclass[12pt,subeqn,a4paper]{article}

\usepackage{amssymb,amsmath,amsfonts,amsthm, amscd, mathrsfs,helvet}
\usepackage{bm}
\usepackage{graphicx,verbatim}
\usepackage{psfrag}
\usepackage[all]{xy}

\def\r2{\sqrt{2}}

\def\bea{\begin{eqnarray} }
\def\eea{\end{eqnarray}}
\def\be{\begin{equation} }
\def\ee{\end{equation}}
\def\nn{\nonumber}

\begin{document}
\newcommand{\nd}[1]{/\hspace{-0.5em} #1}
\begin{titlepage}
\begin{flushright}
{\bf May 2008} \\ 
\end{flushright}
\begin{centering}
\vspace{.2in}
 {\large {\bf A Spin Chain from String Theory}}

\vspace{.3in}

Nick Dorey\\
\vspace{.1 in}
DAMTP, Centre for Mathematical Sciences \\ 
University of Cambridge, Wilberforce Road \\ 
Cambridge CB3 0WA, UK \\
\vspace{.2in}
%
%
\vspace{.4in}
{\bf Abstract} \\
\end{centering}

We study the semiclassical spectrum of bosonic string theory on 
$AdS_{3}\times S^{1}$ in the limit of large $AdS$ angular momentum.  
At leading semiclassical order, this is a subsector of the IIB
superstring on $AdS_{5}\times S^{5}$. 
The theory includes strings
with $K\geq 2$ spikes which approach the boundary in this limit. 
We show that, for all $K$, 
the spectrum of these strings exactly matches that of 
the corresponding operators in the dual gauge theory up to a single
universal prefactor which can be identified with the
cusp anomalous dimension. We propose a precise map between the
dynamics of the spikes and the 
classical $SL(2,\mathbb{R})$ spin chain which arises in
the large-spin limit of ${\cal N}=4$ Super Yang-Mills theory.

\end{titlepage}
\paragraph{}
\section{Introduction}
\paragraph{}
Logarithmic scaling of anomalous 
dimensions with Lorentz spin ($S$) is a characteristic feature of 
composite operators in four-dimensional gauge theory
\cite{GW,K1} (for a recent discussion see \cite{AM1}). Although
initially observed in perturbative gauge theory, the AdS/CFT
correspondence has provided strong evidence that $\log S$ scaling
persists at strong coupling \cite{GKP}. The best understood example is the
anomalous dimension of twist-two operators in the planar limit of
${\cal N}=4$ SUSY Yang-Mills, which has the form, 
\bea \gamma & = &  2\Gamma(\lambda)\,\log(S) \,\,+\,\,O(S^{0})
\nn \eea 
Here $\Gamma$ is a function of the 't Hooft coupling
$\lambda=g^{2}_{\rm YM}N$, known as the cusp anomalous dimension, 
with the following behaviour at weak and
strong coupling respectively\footnote{The emergent integrability of the ${\cal
  N}=4$ theory has subsequently lead to a conjecture \cite{BES} for
$\Gamma(\lambda)$ which should hold for all values of $\lambda$.}, 
\bea 
\Gamma(\lambda)  & = & \frac{\lambda}{4\pi^{2}}\,\,+\,\,O(\lambda^{2})
\qquad{} {\rm for} \,\,\lambda<<1
\nn \\ 
& = & \frac{\sqrt{\lambda}}{2\pi}\,\,+\,\,O(\lambda^{0})\qquad{} 
{\rm for} \,\,  \lambda>>1 \nn \eea 
The strong coupling result was first obtained in \cite{GKP} by
evaluating the classical 
energy of the corresponding state in the dual string theory
on $AdS_{5}\times S^{5}$ in the limit $S\rightarrow\infty$. 
One of the aims of this paper is to extend this analysis to string states
dual to operators of arbitrary twist.
\paragraph{}
In one-loop gauge theory, the large-$S$ spectrum of operators of fixed 
twist has been studied in detail by Belitsky, Gorsky, Korchemsky and 
collaborators (see
\cite{BGK1, BBGK, BGK2} and references therein). 
We will focus on operators of the form,   
\bea 
\hat{O} & \sim & {\rm Tr}_{N}\left[\,\mathcal{D}_{+}^{s_{1}}Z\,
\mathcal{D}_{+}^{s_{2}}Z\,\ldots\, \mathcal{D}_{+}^{s_{J}}Z\,\right] 
\label{sl2} 
\eea
having total Lorentz spin  $S = \sum_{l=1}^{J}\,\,s_{l}$ and twist $J$.  
Here $\mathcal{D}_{+}$ is a covariant derivative with conformal spin
plus one and $Z$ is one of the three complex adjoint scalar fields of the
$\mathcal{N}=4$ theory. For $S>>1$, the resulting one-loop spectrum of
anomalous dimensions lies in a band,     
\paragraph{}
\bea 
\gamma_{\rm min}= \frac{\lambda}{4\pi^{2}}\,\cdot\, 
2\log\left(S\right)\,\, \leq & \gamma_{\rm 1-loop} & 
\leq \,\, \gamma_{\rm max}=\frac{\lambda}{4\pi^{2}}\,\cdot\, 
J\log\left(S\right)
\label{band}
\eea 
More precisely, for each positive integer $K$ with
$2\leq K\leq J$, the large-$S$ theory contains families of states 
labelled by $K-1$ integers $l_{k}\sim S$ with 
$\sum_{k=1}^{K-1}\,l_{k}=S$. The spectrum of these states is,  
\bea 
\gamma_{\rm 1-loop}
[l_{1},l_{2},\ldots, l_{K-1}] & = & \frac{\lambda}{4\pi^{2}}\left( 
K\log\,S \,\,+\,\, H_{K}\left[\frac{l_{1}}{S},\frac{l_{2}}{S},
\ldots,\frac{l_{K-1}}{S}\right]\,\,+\,\, O(1/S)\right) 
\label{spectrum1} \eea 
where the function $H_{K}$ is the Hamiltonian of a 
certain {\em classical} spin chain with $K$
sites. This chain can be thought of as a classical $S\rightarrow
\infty$ limit of the
quantum spin chain which governs the complete one-loop
spectrum of anomalous dimensions (ie for all values of $S$ and $J$), 
in the planar ${\cal N}=4$ theory \cite{MZ, B}. 
The dynamical variables are classical ``spins'' 
$\mathcal{L}^{\pm}_{k}$, $\mathcal{L}^{0}_{k}$, for $k=1,2,\ldots, K$ 
whose Poisson brackets
provide a representation of the $sl(2,\mathbb{R})$ Lie algebra at each
site, 
\bea 
\{ \mathcal{L}^{+}_{k}, \mathcal{L}^{-}_{k'}\} \,=\, 
2i\delta_{kk'} \mathcal{L}^{0}_{k} & \qquad{} & 
\{ \mathcal{L}^{0}_{k}, \mathcal{L}^{\pm}_{k'}\}
\,=\, \pm i\delta_{kk'} 
\mathcal{L}^{\pm}_{k} \label{pb} \eea 
The corresponding quadratic Casimir at each site is appropriate for a
representation of zero spin; 
\bea 
\mathcal{L}^{+}_{k}\,\mathcal{L}^{-}_{k}\,\,+\,\, \left(
\mathcal{L}^{0}_{k}\right)^{2} & = & 0 \label{Cas} \eea 
for $k=1,\ldots, K$. As we review in the next section, 
the classical chain is integrable \cite{FK} 
with a continuous spectrum governed 
by a spectral
curve of genus $K-2$. The discrete spectrum (\ref{spectrum1}) is
obtained by applying appropriate Bohr-Sommerfeld semiclassical 
quantization conditions. 
\paragraph{}
In this paper we will study the corresponding large-$S$ limit
of the dual string theory. Importantly, we are interested in the
$S\rightarrow \infty$ limit with $J$ fixed. Thus the string states we
seek are dual to the states of a spin chain of {\em fixed} length. For 
earlier work relevant to this limit see \cite{Kruc2, Stef, MTT, BGK2, 
Sakai, BE, BES, KS, AM1} and especially \cite{Kruc}. 
Our main result is a calculation the semiclassical string spectrum at
large $S$. We find precise agreement with the gauge theory spectrum 
(\ref{spectrum1}) up to a single overall function of the coupling
which can be identified with the cusp anomalous dimension
$\Gamma(\lambda)$. We will also propose a mechanism whereby 
the gauge theory spin chain emerges as a 
decoupled subsector of semiclassical string theory in the large-$S$
limit. 
The main results are briefly described in the remainder of this
introductory section. 
 \paragraph{}  
For large 't Hooft coupling, operators of the form (\ref{sl2}) are
dual to semiclassical strings moving on an $AdS_{3}\times S^{1}$ 
submanifold of $AdS_{5}\times S^{5}$. Here spin $S$ corresponds to angular
momentum on $AdS_{3}$ and twist $J$ to angular momentum on $S^{1}$. 
Generic solutions of the
equation of motion can be constructed (somewhat implicitly) by the
method of finite gap integration \cite{KZ} (see also 
\cite{Krich, KMMZ, DV1, DV2}). The leading-order semiclassical
spectrum is then obtained by applying the Bohr-Sommerfeld conditions
derived in \cite{DV1, DV2}. 
Solutions are classified by the
number of gaps, $K$, in the spectrum of the auxiliary linear problem. 
These solutions are analogous to classical solutions with $K$
oscillator modes turned on in flat space string theory\footnote{In
  fact this is not just an analogy. In a limit where the equations of
  motion linearize around a pointlike string there is a one-to-one
  correspondence between the mode expansion of the linear problem and
  the gaps of the non-linear one.}. 
\paragraph{}
There are some superficial similarities between the spectrum of
$K$-gap strings and that of the classical spin chain described above. 
In particular both are governed by a hyperelliptic spectral curve and
solutions correspond to linear motion on the Jacobian in both cases. 
However, there are also important differences which reflect the fact that the
string has an infinite number of degrees of freedom while the chain
has a only one degree of freedom on each of its $K$ sites. 
As we review in Section 3, the infinite tower of string modes lead to
essential singularities in the spectral data of the string which are
absent in the corresponding description of the spin chain. A 
related issue is that the string theory curve is only determined
implicitly by the existence of a certain normalised abelian
differential $dp$. The normalisation conditions for $dp$ are
transcendental and generally cannot be solved to give an unconstrained
parametrisation of the curve\footnote{I would like to thank Harry
  Braden for emphasizing this point.}. 
\paragraph{}
In the following we
will consider the large-$S$ limit of the finite gap construction at
fixed $K$ and $J$. 
We will find that large $AdS$ angular momentum leads 
to drastic simplifications which
allow the solution of the period conditions in closed form. The main
result is that the resulting semiclassical string 
spectrum coincides precisely with the
spectrum (\ref{spectrum1}) of the spin chain 
up to the overall coupling dependence. In particular we find,  
\bea 
\gamma_{\rm string}
[l_{1},l_{2},\ldots, l_{K-1}] & = & \frac{\sqrt{\lambda}}{2\pi}\left( 
K\log\,S \,\,+\,\, H_{K}\left[\frac{l_{1}}{S},\frac{l_{2}}{S},
\ldots,\frac{l_{K-1}}{S}\right]\,\,+ \,\,C_{K} \,\,+\,\, O(1/\log S)\right) 
\nn \\ \label{stringspec} \eea  
where $H_{K}$ is the spin chain Hamiltonian, $C_{K}$ 
is an undetermined constant and 
$l_{i}\sim S$ are positive integers such that 
$\sum_{i=1}^{K-1} l_{i}=S$. Worldsheet $\sigma$-model loop corrections
to this semiclassical formula are suppressed by powers of
$1/\sqrt{\lambda}$. The occurence of the first term on the RHS
of (\ref{stringspec}) has been
verified in a previous studies of finite gap solutions \cite{BGK2, Sakai},
the new feature of our result is the detailed agreement with the spin
chain which emerges in the second term which is $O(S^{0})$.  
The results are consistent with the conjecture, 
\bea 
\gamma
[l_{1},l_{2},\ldots, l_{K-1}] & = & \Gamma(\lambda)\left( 
K\log\,S \,\,+\,\, H_{K}\left[\frac{l_{1}}{S},\frac{l_{2}}{S},
\ldots,\frac{l_{K-1}}{S}\right]\,\,+ \,\,C_{K}(\lambda) \right) \,\,+\,\, 
O(1/\log S) 
\nn \eea 
for the exact large-$S$ spectrum where $\Gamma(\lambda)$ is the cusp
anomalous dimension introduced above. A similar
conjecture was made
for the exact spectrum of a
closely related set of operators in large-$N$ QCD in 
\cite{BGK1}\footnote{See Eqn (3.45) in this reference.}. 
\paragraph{}
The $S\rightarrow \infty$ limit considered here is quite 
different from the thermodynamic $J\rightarrow \infty$ limit of the chain  
where the full spectrum is determined by the Asymptotic 
Bethe Ansatz Equations (ABAE) \cite{MS,BS1,BES}. A priori there is no
reason why the ABAE should apply to a spin chain of fixed length. 
On the other hand, it
was argued in \cite{BE}, that the {\em lowest} operator dimension for fixed,
large $S$ is independent of $J$ and can therefore be evaluated in the
$J\rightarrow \infty$ limit using the ABAE\footnote{Indeed this strategy was
used in \cite{BES} to obtain the conjectured exact form of
$\Gamma(\lambda)$.}. The large-$S$ semiclassical spectrum
(\ref{stringspec}) obtained in this paper is also independent of $J$
which suggests that the universality proposed in \cite{BE} may apply
to {\it all} operators with large spin, not just the operator of
lowest dimension.  
\paragraph{}
The exact agreement between the semiclassical spectrum of a 
discrete spin chain and a continuous string initially seems somewhat 
mysterious\footnote{Discreteness here refers to the fact that the spin chain
  lives on a spatial lattice. A related mystery for the case of the magnon
  dispersion relation is resolved in \cite{HM}.}. 
In the final part of the paper, we will propose a
  precise account of how the classical spins naturally emerge from
  the string at large $S$. 
The key phenomenon is already visible in 
in the rotating folded string studied in \cite{GKP}. Logarithmic
  scaling of the form $\Delta-S\simeq
  \sqrt{\lambda}/{2\pi}\,\cdot\,2\log S$ naturally arises 
arises when the two folds of the string approach the boundary of
  $AdS_{3}$. It is natural to expect that the finite gap
  solutions studied in this paper correspond to configurations with
  $K$ spikes which approach the boundary as $S\rightarrow \infty$
  giving the scaling $\Delta-S\simeq
  \sqrt{\lambda}/{2\pi}\,\cdot\,K\log S$. Rigidly rotating solutions of this
  type were constructed in \cite{Kruc}. 
\paragraph{}
In static conformal gauge, the string $\sigma$-model coincides with
  the the $SL(2,\mathbb{R})$ 
Principal Chiral Model. The dynamical
  variable is the Noether current 
$j_{\pm}(\sigma,\tau)=g^{-1}\partial_{\pm}g$ 
with $g(\sigma,\tau)\in SL(2,\mathbb{R})$ corresponding to right
  multiplication in the group.  In the limit $S\rightarrow \infty$ we
  will argue that the
  corresponding charge density 
becomes $\delta$-function localised at the positions of 
the K spikes. This localisation leads to a natural proposal for
  spin degrees of freedom, 
 \bea
 {L}^{A}_{k}  & = & \lim_{S\rightarrow \infty} \,\,
\left[\frac{\sqrt{\lambda}}{8\pi}
\int_{\mu_{k}}^{\mu_{k+1}}\,d\sigma\, j^{A}_{\tau}(\sigma,\tau)\right]
\label{def} \eea        
for $k=1,2,\ldots K$ where the index $A=0,1,2$ runs over the
  generators of $SL(2,\mathbb{R})$. Here, the $K$'th spike is located at
  $\sigma=\sigma_{k}$ and $\mu_{k}$ is are arbitrary points on the
  string with  
$\mu_{k}<\sigma_{k}<\mu_{k+1}$. We propose that the variables defined
  in (\ref{def}) are related
  to the spins introduced above as, $\mathcal{L}^{0}_{k}=
L^{0}_{k}$ and $i\mathcal{L}^{\pm}_{k}=L^{1}_{k}\pm iL^{2}_{k}$. 
In particular one may then verify the Poisson brackets (\ref{pb}) and
  the quadratic Casimir relation (\ref{Cas}). 
With this identification it can be
  shown that the monodromy matrix of the string reduces 
to that of the classical 
spin chain as $S\rightarrow \infty$. A related correspondence between
  the dynamics of spikes and the spin chain was suggested in
  \cite{Kruc}. The emergence of a spin chain from a limit of string
  theory was also discussed in \cite{Kazakov}. 
\paragraph{}
The remainder of the paper is organised as follows. In Section 2 we
provide a brief review of the semiclassical spin chain which arises in
one-loop gauge theory. In Section 3 we introduce the finite gap
construction of string solutions and the corresponding spectral curve 
$\Sigma$. In Section 4 we study the $S\rightarrow \infty$ limit for
generic solutions and show that it corresponds to a particular
degeneration of the spectral curve. In Section 5, we solve the 
period conditions for the differential $dp$ in the degenerate limit
and find the semiclassical spectrum of the model. In Section 6 we give
an interpretation of our results in terms of spikey strings and
propose a precise map between spikes and strings. Finally the results
are discussed in Section 7.  

\section{The gauge theory spin chain}
\paragraph{}
We consider the one-loop anomalous dimensions of operators in the  
non-compact rank one subsector of planar $\mathcal{N}=4$ SUSY Yang-Mills 
(also known as the $sl(2)$ sector). Generic single-trace operators in
this sector are labelled by their Lorentz spin $S$ and $U(1)_{R}$
charge $J$ and have the form (\ref{sl2}).  The classical dimension 
of each operator is $\Delta_{0}=J+S$ and its twist (classical
dimension minus spin) is therefore equal to $J$. 
\paragraph{}
The one-loop anomalous dimensions of operators in the $sl(2)$ sector
are determined by the eigenvalues of the Hamiltonian of the 
Heisenberg XXX$_{-\frac{1}{2}}$ spin chain with $J$ sites. Each site
of this chain 
carries a representation of $SL(2,\mathbb{R})$ with quadratic Casimir
equal to minus one half. Our discussion of the chain in this section
follows that of \cite{BGK1,BGK2} (See in particular Section 2.2 of
\cite{BGK2}). Here we will focus on the large-spin limit of
the chain: $S\rightarrow \infty$ with $J$ fixed. This is a
effectively a semiclassical limit where $1/S$ plays the role of
Planck's constant $\hbar$ \cite{BGK1,BGK2}. In this limit the 
quantum spins are replaced by the classical variables 
$\mathcal{L}^{\pm}_{k}$, $\mathcal{L}^{0}_{k}$, for $k=1,2,\ldots, J$, 
introduced above. The commutators of spin operators are
replaced by the Poisson brackets (\ref{pb}) of these classical spins. 
As the quadratic Casimir
equal to $-1/2$ is negligable in the $S\rightarrow \infty$ limit, 
the classical spins at
each site obey the relation (\ref{Cas}) up to $1/S$ corrections. 
We will restrict our attention to states obeying the highest weight
condition, 
\bea 
\sum_{k=1}^{J} \mathcal{L}^{\pm}_{k} & = &  0  
\label{hw} \eea 
\paragraph{}
Integrability of the classical spin chain starts from the existence 
of a Lax matrix, 
\bea 
\mathbb{L}_{k}(u) & = & \left(\begin{array}{cc} u+
i\mathcal{L}^{0}_{k} & i\mathcal{L}^{+}_{k} \\ i\mathcal{L}^{-}_{k}
& u-i\mathcal{L}^{0}_{k} \end{array}\right) 
\nn \eea
where $u\in \mathbb{C}$ is a spectral parameter.  
A tower of conserved quantities are obtained by constructing the monodromy, 
\bea 
t_{J}(u) & = & {\rm
  tr}_{2}\left[\mathbb{L}_{1}(u)\mathbb{L}_{2}(u)
\ldots\mathbb{L}_{J}(u)\right]   \nn \\ 
& = & 2u^{J}\,+\, q_{2}u^{J-2} \,+
\,\ldots\,+\,q_{J-1}u\, +\, q_{J} \label{monod3} \eea
At large-$S$ we find 
$q_{2}=-S^{2}$ up to corrections of order $1/S$. 
One may check starting from the Poisson brackets (\ref{pb}) that the
conserved charges, $q_{j}$, $j=2,3,\ldots J$ are in involution: 
$\{q_{j}, \,\, q_{k}\}=0$ $\forall\, j,\,k$. Taking into account the
highest-weight constraint (\ref{hw}), this is a sufficient number of
conserved quantities for complete integrability of the chain. 
\paragraph{}
The one-loop spectrum of operator
dimensions at large-$S$ is determined from the semiclassical spectrum of
the spin chain. It has different branches, labelled by an integer
$K\leq J$, corresponding to the highest
non-zero conserved charge \cite{BGK2}, 
\bea 
q_{K}\neq 0 & \qquad{} & q_{j}=0\qquad{}\forall\,\,j>K \nn \eea 
The one-loop anomalous dimensions are given as, 
\bea \gamma_{\rm one-loop} \,\, = \,\, \Delta-J-S & = & 
\frac{\lambda}{8\pi^{2}}\log\left(q_{K}\right)\,\,+\,\,O(1/S)
\label{spec}
\eea 
We call the branch with $K=J$ the highest sector. For each  
$K<J$ there is also a sector of states 
isomorphic to the highest sector of a shorter chain with only $K$
sites. 
In the limit of large-$S$, the conserved charge $q_{j}$ scales as
$S^{j}$ for $j=2,\ldots,K$. Hence (\ref{spec}) exhibits the expected
logarithmic scaling with $S$. 
In the following it will be useful
to introduced rescaled charges $\hat{q}_{j}$, such that  
$q_{j}=S^{j}\,\hat{q}_{j}$. In particular $\hat{q}_{2}=-1$ up to
corrections of order $1/S$.  
\paragraph{}
At the classical level, the conserved charges $\hat{q}_{j}$ vary
continuously. The discrete spectrum described in the Introduction
arises from imposing appropriate Bohr-Sommerfeld quantisation
conditions. To describe these we introduce the spectral curve of the
spin chain,  
\bea 
\Gamma_{K} \,:\qquad{} y^{2} & = & \prod_{l=1}^{K-2}\, (x-x_{l}) \nn
\\ 
& = & x^{2K}\left[1- 
\frac{1}{4}\hat{\mathbb{P}}_{K}\left(\frac{1}{x}\right)^{2}\right] \nn \eea 
with, 
\bea 
\hat{\mathbb{P}}_{K}
\left(\frac{1}{{x}}\right) & = & 2\,\,-\,\,\frac{1}
{{x}^{2}}\,\,+\,\, \frac{\hat{q}_{3}}
{{x}^{3}}\,\,+\,\,\ldots\,\,+\,\, \frac{\hat{q}_{K}}
{{x}^{K}} \nn \eea 
which is a hyperelliptic Riemann surface of genus $K-2$. The rescaled
spectral parameter $x=u/S$ is held fixed as $S\rightarrow \infty$ 
and the rescaled conserved charges 
$\hat{q}_{j}$, $j=3,\ldots,K$ correspond to moduli of the curve. 
Notice that the curve
is highly non-generic in that the positions of the $2K-2$ branch points
$x_{l}$ are determined in terms of $K-2$ parameters $\hat{q}_{j}$. 
\begin{figure}
\centering
\psfrag{x}{\footnotesize{$x$}}
\psfrag{m1}{\footnotesize{$-1$}}
\psfrag{p1}{\footnotesize{$+1$}}
\psfrag{C1m}{\footnotesize{$C_{1}^{-}$}}
\psfrag{C1p}{\footnotesize{$C_{1}^{+}$}}
\psfrag{C1k}{\footnotesize{$C_{K/2}^{-}$}}
\psfrag{C1j}{\footnotesize{$C_{K/2}^{+}$}}
\psfrag{a}{\footnotesize{$x_{2K-2}$}}
\psfrag{b}{\footnotesize{$x_{2K-1}$}}
\psfrag{c}{\footnotesize{$x_{4}$}}
\psfrag{d}{\footnotesize{$x_{3}$}}
\psfrag{e}{\footnotesize{$x_{2}$}} 
\psfrag{f}{\footnotesize{$x_{1}$}}
\psfrag{A1}{\footnotesize{$\alpha_{1}$}}
\psfrag{A2}{\footnotesize{$\alpha_{2}$}}
\psfrag{A3}{\footnotesize{$\alpha_{K-1}$}}
\psfrag{h}{\footnotesize{$a^{(K-1)}_{+}$}}
\includegraphics[width=100mm]{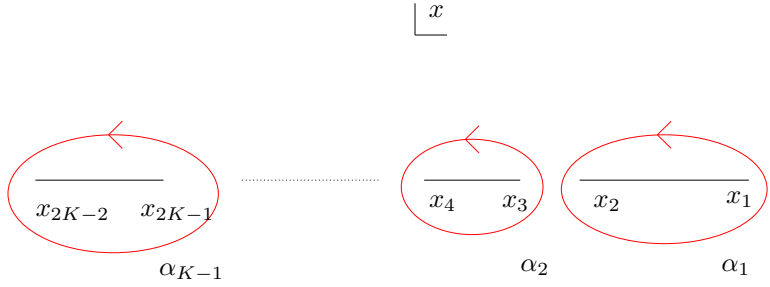}
\caption{The cut $x$-plane corresponding to the curve $\Gamma_{K}$}
\label{Sfig1b}
\end{figure}
The curve $\Gamma_{K}$ corresponds to a double cover of the $x$ plane as
shown in Figure 1. We also define $K-1$ one-cycles $\alpha_{j}$,
$j=1,\ldots, J-1$ as shown in the Figure\footnote{To state the main
  results of \cite{BGK1, BGK2}, we will not need a
to introduce a full basis of cycles on $\Gamma_{K}$.}.  
\paragraph{}
The Bohr-Sommerfeld conditions are expressed in terms of a certain
meromorphic differential on $\Gamma_{K}$, 
\bea d\hat{p} & = & -i\,\frac{d{x}}{{x}^{2}}\,\,
\frac{\hat{\mathbb{P}}'_{K}\left(\frac{1}{{x}}\right)}
{ \sqrt{\hat{\mathbb{P}}_{K}
\left(\frac{1}{{x}}\right)^{2}\,\,-\,\,4}} 
\label{dhp1} \eea  
and they read, 
\bea 
-\frac{1}{2\pi i} \,\oint_{\alpha_{j}}\,x\,d\hat{p} & = &
\frac{l_{j}}{S} \qquad{} l_{j}\in \mathbb{Z}^{+} \label{BS1} \eea
for $j=1,2,\ldots, K-1$. The integers $l_{j}$ which label the states
in the spectrum obey the conditions, 
\bea 
\sum_{j=1}^{K-1} \,l_{j}\,=\,S, & \qquad{} & \sum_{j=1}^{K-1} \,j
l_{j}\,=\, 0 \,\,{\rm mod}\,\,K \label{BS2} \eea 
The first equality is related to the fact that the integers $l_{j}$
count numbers of Bethe roots associated with each cut.  
Each root carries one unit of spin and thus the total number of roots is
equal to $S$. The second condition is imposed by the cyclicity of the
trace and
corresponds to a vanishing of the total momentum of the chain. 
\paragraph{}
In order for a
leading order 
semiclassical approach for any quantum mechanical problem to be valid
it is necessary that the quantum numbers are large. Thus we must take
$l_{j}\sim O(S)$ as $S\rightarrow \infty$ for each $j$. Then both
sides of Eqn (\ref{BS1}) scale like $S^{0}$.  
The $K-2$ independent equations (\ref{BS1}) determine the charges, 
\bea \hat{q}_{j} & = &
\hat{q}_{j}\left[\frac{l_{1}}{S},\frac{l_{2}}{S},\ldots,\frac{l_{K-1}}{S}
\right] 
\nn \eea 
$j=3,\ldots, K$. Finally the spectrum of one-loop anomalous dimensions
is given as, 
\bea
\gamma[l_{1},\ldots,l_{K-1}] & = & 
\frac{\lambda}{4\pi^{2}}\left( K\log S\,\,+\,\,
H_{K}\left[\frac{l_{1}}{S},\frac{l_{2}}{S},
\ldots, \frac{l_{K-1}}{S}\right]\,\,+\,\,O(1/S)\right)\label{spectrum2} \eea
where $H_{K}=\log \hat{q}_{K}$. 

\section{Semiclassical string theory}
\paragraph{}
At large 't Hooft coupling $\sqrt{\lambda} >> 1$, gauge theory 
operators of the $sl(2)$ sector are dual to semiclassical strings
moving on $AdS_{3} \times S^{1}$. The $U(1)$ R-charge $J$ corresponds
to momentum in the $S^{1}$ direction and the conformal spin $S$
corresponds to angular momentum in $AdS_{3}$. We introduce string
worldsheet coordinates $\sigma \sim \sigma + 2\pi$ and $\tau$ and 
the corresponding lightcone coordinates $\sigma_{\pm}=(\tau\pm
\sigma)/2$ and we define lightcone derivatives
$\partial_{\pm}=\partial_{\tau} \pm \partial_{\sigma}$. The space-time
coordinates correspond to fields on the string worldsheet: we
introduce $\phi(\sigma,\tau)\in S^{1}$ and parametrize $AdS_{3}$ with
a group-valued field $g(\sigma,\tau)\in SL(2,\mathbb{R})\simeq
AdS_{3}$. The $SL(2,\mathbb{R})_{R}\times SL(2,\mathbb{R})_{L}$
isometries of $AdS_{3}$ correspond to left and right group
multiplication. The Noether current corresponding to right
multiplication is $j_{\pm}=g^{-1}\partial_{\pm}g$. Following \cite{KZ}, we
work in static, conformal gauge with a flat worldsheet metric and set, 
\bea \phi(\sigma,\tau) & = & \frac{J}{\sqrt{\lambda}}\,\,\tau \nn \eea 
In this gauge, the string action becomes that of the $SL(2,\mathbb{R})$ 
Principal Chiral model, 
\bea 
S_{\sigma} & = & \frac{\sqrt{\lambda}}{4\pi}\, \int_{0}^{2\pi}
\,d\sigma\, \frac{1}{2}{\rm tr}_{2}\left[j_{+}j_{-}\right] \nn \eea 
String motion is also subject to the Virasoro constraint, 
\bea 
\frac{1}{2}{\rm tr}_{2}\left[j_{\pm}^{2}\right] & = &
\frac{J^{2}}{\lambda}
\nn \eea 
 \paragraph{}
Classical integrability of string theory on $AdS_{3}\times S^{1}$
follows from the construction of the monodromy matrix \cite{BPR}, 
\bea 
\Omega\left[x;\tau \right] & = &
\,\,\mathcal{P}\,\,\exp\left[\frac{1}{2}
\int_{0}^{2\pi}\,d\sigma \,\, \left( \frac{j_{+}}{x-1} \,\,+\,\,
\frac{j_{-}}{x+1}\right )\right] \,\,\,\,\in\,\,\,\, SL(2,\mathbb{R})
\nn \eea 
whose eigenvalues $w_{\pm}=\exp(\pm i\,p(x))$ are
$\tau$-independent for all values of the spectral parameter $x$. It is
convenient to consider the analytic continuation of the monodromy
matrix $\Omega[x;\tau]$ and of the quasi-momentum $p(x)$ to complex
values of $x$. In this case $\Omega$ will take values in
$SL(2,\mathbb{C})$ and appropriate reality conditions must be imposed
to recover the physical case.   
\paragraph{}
The eigenvalues $w_{\pm}(x)$ are two branches of an analytic
function defined on the spectral curve, 
\bea 
\Sigma_{\Omega}\,\,: \qquad{} \qquad{} 
{\rm det}\left(w \mathbb{I} \,\,-\,\,\Omega[x;\tau]
\right)\,\,=\,\, w^{2}\,\,-\,\,2\cos p(x) \, w\,\,+\,\,1  & =
& 0 \qquad{} w, \,\,x\in \mathbb{C} \nn \eea 
This curve corresponds to a double cover of the complex $x$-plane with
branch points at the simple zeros of the discriminant
$D=4\sin^{2}p(x)$. In addition the monodromy matrix defined above is
singular at the points above $x=\pm 1$. Using the Virasoro constraint,
one may show that, $p(x)$ has a simple poles at these points,  
\bea 
p(x) & \sim & \frac{\pi J}{\sqrt{\lambda}}\frac{1}{(x\pm 1)^{2}} \,+\, 
O\left((x\pm 1)^{0}\right) \label{simp} \eea
as $x\rightarrow \mp 1$. Hence
the discriminant $D$ has essential singularities at $x=\pm 1$ and $D$
must therefore have an infinite number of zeros which
accumulate at these points. Formally we may represent the discriminant
as a product over its zeros and write the spectral curve as 
\bea 
\Sigma_{\Omega}\,\,: \qquad{} \qquad{}  {y}^{2}_{\Omega}\,\,=
\,\,4\sin^{2}p(x) & = & \prod_{j=1}^{\infty} \left(x-x_{i}\right) \nn
\eea 
For generic solutions the points $x=x_{i}$ are distinct and the curve 
${\Sigma}_{\Omega}$ has infinite genus. 
\paragraph{}
In order to make progress it is necessary to focus on 
solutions for which the discriminant has only a finite number $2K$ of 
simple zeros and the spectral curve ${\Sigma}_{\Omega}$ has finite genus. 
The infinite number of additional zeros of the discriminant $D$ must
then have multiplicity two or higher. 
These are known as {\rm finite gap solutions}\footnote{Strictly
  speaking these are not generic solutions of the string equations of
  motion. However, as $K$ can be
arbitrarily large, it is reasonable to expect that generic solutions
could be obtained by an appropriate $K\rightarrow \infty$ limit.}.  
In this case, $dp$ is a
meromorphic differential on the hyperelliptic curve, 
\bea 
\Sigma\,\,:\qquad{} y^{2} & = & \prod_{i=1}^{2K} \left(x-x_{i}\right)
\nn \eea 
of genus $g=K-1$ which is obtained by removing the double points of 
$\hat{\Sigma}$ (see \cite{DV1}). For ease of presentation 
we will consider only even values of $K$, the generalisation to odd
values is straightforward. 
\paragraph{}
In the following we will focus on curves where all the branch points
lie on the real axis and outside the interval $[-1,+1]$. This
corresponds to string solutions where only classical oscillator modes
which carry positive spin are activated. In the dual gauge theory
these solutions are believed to correspond to operators of the form 
(\ref{sl2}) where only the covariant derivative 
$\mathcal{D}_{+}$, which carries positive spin, appears \cite{KZ, min}. 
We label the branch points of the curve according to,  
\bea 
\Sigma\,\,:\qquad{} y^{2} & = & \left(x-b_{+}\right)  
\left(x-b_{-}\right) \prod_{i=1}^{K-1} 
\left(x-a^{(i)}_{+}\right)\left(x-a^{(i)}_{-}\right)
\label{curve} \eea
with the ordering, 
\bea 
a_{-}^{(K-1)}\,\,\leq \,\,a_{-}^{(K-2)} & \ldots &\leq
a_{-}^{(1)}\,\,\leq b_{-}\,\,\leq -1  \nn \\ 
a_{+}^{(K-1)}\,\,\geq \,\,a_{+}^{(K-2)} & \ldots &\geq
a_{+}^{(1)}\,\,\geq b_{+}\,\,\geq +1 \nn \eea 
The branch points are joined in pairs by cuts $C^{\pm}_{I}$, 
$I=1,2,\ldots,K/2$
as shown in Figure 2.  
We also define a standard basis of one-cycles,
$\mathcal{A}^{\pm}_{I}$, $\mathcal{B}^{\pm}_{I}$. Here
$\mathcal{A}^{\pm}_{I}$ encircles the cut $C^{\pm}_{I}$ on the upper
sheet in an anti-clockwise direction and $\mathcal{B}^{\pm}_{I}$ runs
from the point at infinity on the upper sheet to the point at infinity
on the lower sheet passing through the cut  $C^{\pm}_{I}$, as shown in
Figure 3. For any $x_{0}\in \mathbb{C}$, 
we will sometimes use the notation $x_{0}^{\pm}$ to denote the
two points on $\Sigma$ where $x=x_{0}$.  
\begin{figure}
\centering
\psfrag{x}{\footnotesize{$x$}}
\psfrag{m1}{\footnotesize{$-1$}}
\psfrag{p1}{\footnotesize{$+1$}}
\psfrag{C1m}{\footnotesize{$C_{1}^{-}$}}
\psfrag{C1p}{\footnotesize{$C_{1}^{+}$}}
\psfrag{C1k}{\footnotesize{$C_{K/2}^{-}$}}
\psfrag{C1j}{\footnotesize{$C_{K/2}^{+}$}}
\psfrag{a}{\footnotesize{$a^{(K-1)}_{-}$}}
\psfrag{b}{\footnotesize{$a^{(K-2)}_{-}$}}
\psfrag{c}{\footnotesize{$a^{(1)}_{-}$}}
\psfrag{d}{\footnotesize{$b_{-}$}}
\psfrag{e}{\footnotesize{$b_{+}$}} 
\psfrag{f}{\footnotesize{$a^{(1)}_{+}$}}
\psfrag{g}{\footnotesize{$a^{(K-2)}_{+}$}}
\psfrag{h}{\footnotesize{$a^{(K-1)}_{+}$}}
\includegraphics[width=100mm]{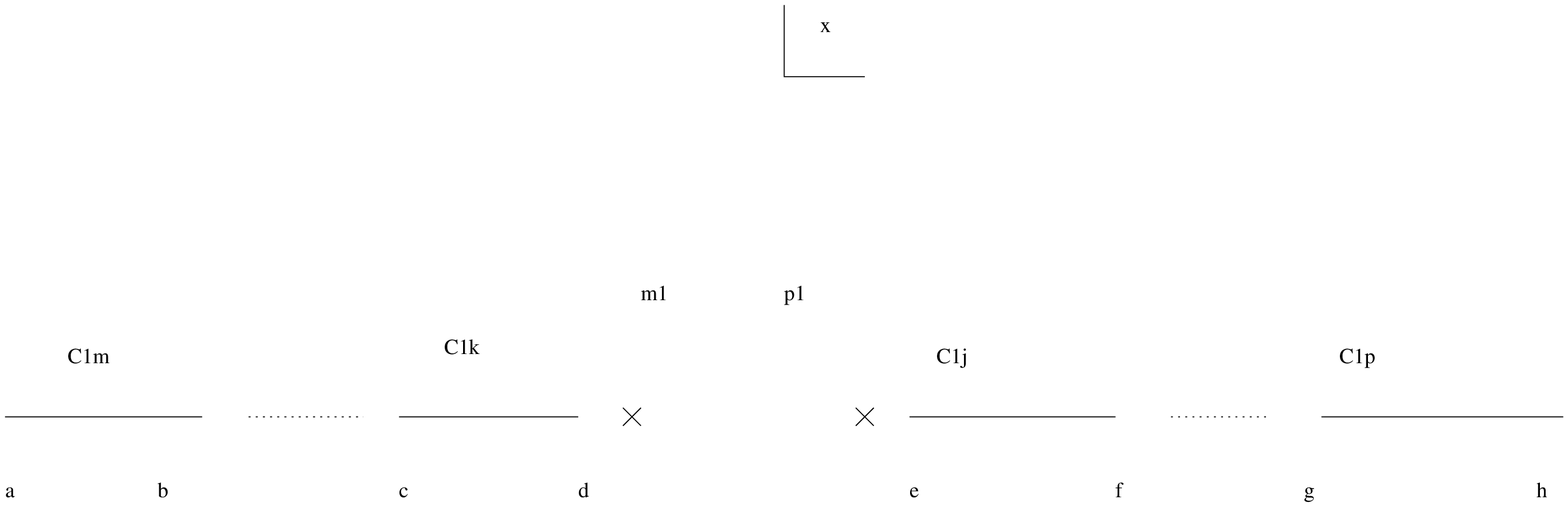}
\caption{The cut $x$-plane corresponding to the curve $\Sigma$}
\label{Sfig1}
\end{figure}
\begin{figure}
\centering
\psfrag{AI}{\footnotesize{$\mathcal{A}^{\pm}_{I}$}}
\psfrag{BI}{\footnotesize{$\mathcal{B}^{\pm}_{I}$}}
\psfrag{CI}{\footnotesize{$C^{\pm}_{I}$}}
\psfrag{Ip}{\footnotesize{$\infty^{+}$}}
\psfrag{Im}{\footnotesize{$\infty^{-}$}}
\includegraphics[width=100mm]{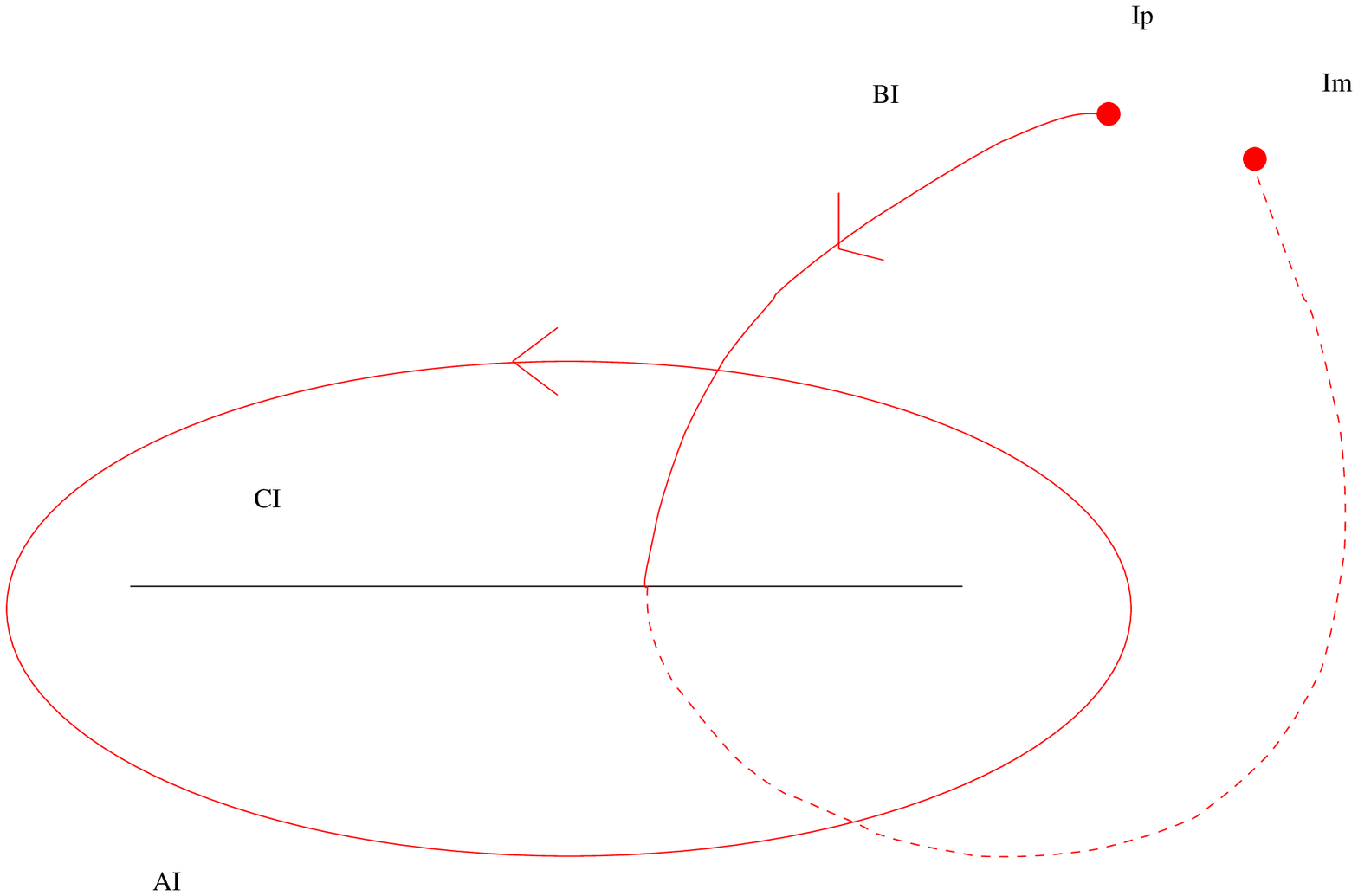}
\caption{The cycles on $\Sigma$. The
  index $I$ runs from $1$ to $K/2$.}
\label{Sfig2}
\end{figure}
\paragraph{}
The quasi-momentum $p(x)$ gives rise a meromorphic abelian
differential $dp$ on $\Sigma$. From (\ref{simp}) we see that the 
differential has second-order poles at the points above $x=+1$ and
$x=-1$ on $\Sigma$. On the top sheet we have, 
\bea 
dp & \longrightarrow & -\frac{\pi J}{\sqrt{\lambda}}\,\, \frac{dx}{(x\pm
    1)^{2}}\,\, + \,\, O\left((x\pm 1)^{0}\right)  
\label{poles} 
\eea 
as $x\rightarrow \mp 1$. There are also two second-order poles at the
points $x=\mp 1$ on the lower sheet related by the involution 
$dp\rightarrow -dp$. The value of the Noether charges $\Delta$ and $S$
is encoded in the asymptotic behaviour of $dp$ near the points $x=0$
and $x=\infty$ on the top sheet, 
\bea
dp & \longrightarrow & -\frac{2\pi}{\sqrt{\lambda}}\,\,
\left(\Delta\,+\,S \right)\,\,\frac{dx}{x^{2}} \qquad{} {\rm as}\,\,\,\, 
x\rightarrow \infty \label{asymp1} \\ 
dp & \longrightarrow & -\frac{2\pi}{\sqrt{\lambda}}\,\,
\left(\Delta\,-\,S\right)\,\,dx \qquad{} {\rm
  as}\,\,\,\, 
x\rightarrow 0 \label{asymp2} \eea
For a valid semiclassical description, the conserved charges $J$, $S$ and
$\Delta$ should all be $O(\sqrt{\lambda})$ with $\sqrt{\lambda}>>1$.  
\paragraph{}
In addition to the above relations, 
$dp$ must obey $2K$ normalisation conditions, 
\bea 
\oint_{\mathcal{A}^{\pm}_{I}} \,dp\,\,=\,\,0   & \qquad{} & 
\oint_{\mathcal{B}^{\pm}_{I}} \,dp \,\,=\,\,2\pi n^{\pm}_{I} 
\label{norm} \eea 
with $I=1,2,\ldots, K/2$. The integers $n^{\pm}_{I}$ correspond to the
mode numbers of the string. In the following we 
will assign the mode numbers so as to
pick out the $K$ lowest modes of the string which carry positive
angular momentum, 
including both left and
right movers. This is accomplished by setting $n^{\pm}_{I}=\pm I$ for
$I=1,\ldots, K/2$. 
\paragraph{}
To find the spectrum of classical string solutions
we must first construct the 
meromorphic differential $dp$ with the specified pole
behaviour (\ref{poles}). The most general possible such differential
has the form, 
\bea 
dp  & = & dp_{1}\,\,+\,\,dp_{2}\,\,=\,\,-\frac{dx}{y}\left[f(x)\,\,+\,\,
g(x)\right]  \label{ansatz} \\ 
f(x) & = & \sum_{\ell=0}^{K-2}\,\,C_{\ell}\,x^{\ell} \nn \\
g(x) & = & \frac{\pi J}{\sqrt{\lambda}}\left[
\frac{y_{+}}{(x-1)^{2}}\,\,+\,\,\frac{y_{-}}{(x+1)^{2}} 
+\frac{y'_{+}}{(x-1)}\,\,-\,\,\frac{y'_{-}}{(x+1)}\right] \nn \eea 
with $y_{\pm}=y(\pm 1)$ and 
\bea 
y'_{\pm} & = & \left. \frac{dy}{dx}\right|_{x=\pm 1} \nn \eea  
Here the second term $dp_{2}$ is a particular differential with the
required poles and the first term $dp_{1}$ is a general holomorphic
differential on $\Sigma$. The resulting curve $\Sigma$ and
differential $dp$ depend on
$3K-1$ undetermined parameters $\{b_{\pm}, a^{(i)}_{\pm}, C_{\ell}\}$
with $i=1,\ldots, K-1$, $\ell=0,1,\ldots,K-2$. We then obtain $2K$
constraints on these parameters from the normalisation equations
(\ref{norm}), leaving us with a $K-1$ dimensional moduli space of
solutions \cite{KZ}. As mentioned above, a significant difficulty with this
approach is that the normalisation conditions are transcendental and
cannot be solved in closed form. 
\paragraph{} 
A convenient parametrisation for the moduli space is given in terms of
the $K$ {\em filling fractions}, 
\bea 
\mathcal{S}^{\pm}_{I} & = & \frac{1}{2\pi i}\,\cdot\,
\frac{\sqrt{\lambda}}{4\pi}\,\,\oint_{\mathcal{A}^{\pm}_{I}}\,
\left(x\,+\,\frac{1}{x}\right)\,dp \nn \eea 
with $I=1,\ldots,K/2$, subject to the level matching constraint, 
\bea 
\sum_{I=1}^{K/2}\,\,n^{+}_{I}\mathcal{S}^{+}_{I}\,\,+\,\,
n^{-}_{I}\mathcal{S}^{-}_{I} & = & 0  \nn \eea
Here the total $AdS$ angular momentum is given as 
\bea 
S & = & \sum_{I=1}^{K/2}\,\mathcal{S}^{+}_{I}\,\,+\,\,
\mathcal{S}^{-}_{I} \nn \eea
and is regarded as one of the moduli of the solution. The significance
of the filling fractions is that they constitute a set of 
normalised action variables for the string\footnote{
The symplectic structure of the string was analysed in detail for the case of
  strings on $S^{3}\times \mathbb{R}$ in \cite{DV1,DV2}. The resulting
  string $\sigma$-model was an $SU(2)$ principal chiral model (PCM). In the
  context of the finite gap construction one works with a complexified
  Lax connection and results for the $SU(2)$ 
and $SL(2,\mathbb{R})$ PCMs differ only at the level of reality
  conditions which do not affect the conclusion that the
filling fractions are the canonical action variables of the string.}. 
They are canonically
conjugate to angles $\varphi_{I}\in [0,2\pi]$ living on the Jacobian
torus $\mathcal{J}(\Sigma)$. Evolution of the string solution in both
worldsheet coordinates, $\sigma$ and $\tau$, corresponds to linear
motion of these angles \cite{DV1}.  
\paragraph{}
The constraints described above uniquely determine $(\Sigma,dp)$ for
given values of $\mathcal{S}^{\pm}_{I}$, and 
one may then extract the 
string energy from the asymptotics (\ref{asymp1},\ref{asymp2}) which imply,  
\bea 
\Delta\,+\,S & = & -\frac{\sqrt{\lambda}}{2\pi}\, C_{K-2} \nn \\ 
\Delta\,-\,S & = & -\frac{\sqrt{\lambda}}{2\pi}\, \frac{C_{0}}{y(0)}
\,\,+ \,\, \frac{J}{2y(0)}\left(y_{+}+y_{-}-y'_{+}-y'_{-}\right) \nn 
\eea
In this way, one obtains a set of transcendental equations which determine
the string energy as a function of the filling fractions,  
\bea
\Delta & = & \Delta\left[\mathcal{S}^{+}_{1},\mathcal{S}^{-}_{1}, \ldots, 
\mathcal{S}^{+}_{K/2},\mathcal{S}^{-}_{K/2}\right] \nn \eea 
Finally the leading order semiclassical spectrum of the string is
obtained by imposing the Bohr-Sommerfeld conditions which impose the
integrality of the filling fractions \cite{DV1, DV2}: $\mathcal{S}^{\pm}_{I}\in
\mathbb{Z}$, $I=1,2,\ldots, K/2$. For uniform validity of the semi-classical
approach we should focus on states where 
$\mathcal{S}^{\pm}_{I}\sim \sqrt{\lambda}$ for each $I$. Higher-loop
corrections in the string $\sigma$-model are then supressed by powers of
$1/\sqrt{\lambda}$.  

\section{The large-$S$ limit}
\paragraph{}
In this section we will take an $S\rightarrow \infty$ limit with fixed
$J$ for the generic $K$-gap solution. The 't Hooft coupling
$\lambda>>1$ is also held fixed in the limit. In the genus one ($K=2$) 
case this limit has been studied in \cite{BGK2}. At the level of the curve
(\ref{curve}),  
the ``outer'' branch points $a^{(1)}_{\pm}$ of the $K=2$ curve scale linearly
with $S$ approaching infinity in the large-$S$ limit, while the inner
branch points $b_{\pm}$ approach the singular points of $dp$ at $x=\pm
1$. For $K>2$ we will take a similar limit where the $2K-2$ branch points 
$a^{(i)}_{\pm}$ will all scale linearly with the spin. 
To impliment this we set,  
\bea
a^{(i)}_{\pm} & = & \rho \tilde{a}^{(i)}_{\pm} \nn \eea 
for $i=1,2,\ldots,K-1$ and take the limit $\rho\rightarrow \infty$
with $\tilde{a}^{(i)}_{\pm}$ held fixed. The remaining branch points,
$b_{\pm}$, are treated as $O(\rho^{0})$. Eventually we will see that
$S\sim \rho$ and also that we 
are forced to take  
$b_{\pm}\rightarrow 1$ as $\rho \rightarrow\infty$ as in the genus one case
of \cite{BGK2}. A related limit of the $K$-gap solution was studied in
\cite{Sakai}. For convenience we will also set $b_{+}=-b_{-}=b\geq
1$ although the same results are obtained without this condition.  
\paragraph{}
Our main concern is to analyse the limiting behaviour of the equations
(\ref{curve}, \ref{norm}, \ref{ansatz}) which define the pair 
$(\Sigma,dp)$. The limit has a convenient description in terms of a 
{\em degeneration} of the spectral curve $\Sigma$. The relevant
degeneration is one where the closed cycle $\hat{\mathcal{B}}=
\mathcal{B}^{+}_{K/2}-\mathcal{B}^{-}_{K/2}$ on $\Sigma$ pinches at
two points as shown in Figure 4. The result is that the curve
$\Sigma$, which has genus $K-1$, 
factorizes into two components,  
\bea 
\Sigma & \longrightarrow & \tilde{\Sigma}_{1}\,\,\cup\,\,
\tilde{\Sigma}_{2} \label{degen} \eea 
where $\tilde{\Sigma}_{1}$ is a curve of genus $K-2$
and $\tilde{\Sigma}_{2}$ is a curve of genus zero. There are two
additional marked points on each component where the two curves touch.   
The degeneration of the curve is determined by the condition that the
differential $dp$ has a good limit as $\rho\rightarrow \infty$. 
The main point is that, as $\rho\rightarrow \infty$ 
the normalisation conditions for the
differential $dp$ on $\Sigma$ reduce to conditions on two meromorphic
differentials $d\tilde{p}_{1}$ and $d\tilde{p}_{2}$ defined on the
curves $\tilde{\Sigma}_{1}$ and $\tilde{\Sigma}_{2}$ respectively.  
\begin{figure}
\centering
\psfrag{S1}{\footnotesize{$\Sigma$}}
\psfrag{S2}{\footnotesize{$\tilde{\Sigma}_{2}$}}
\psfrag{B}{\footnotesize{$\hat{\mathcal{B}}$}} 
\psfrag{A2}{\footnotesize{$\hat{\mathcal{A}}_{2}$}}
\psfrag{Ip}{\footnotesize{$\infty^{+}$}}
\psfrag{Im}{\footnotesize{$\infty^{-}$}}
\includegraphics[width=100mm]{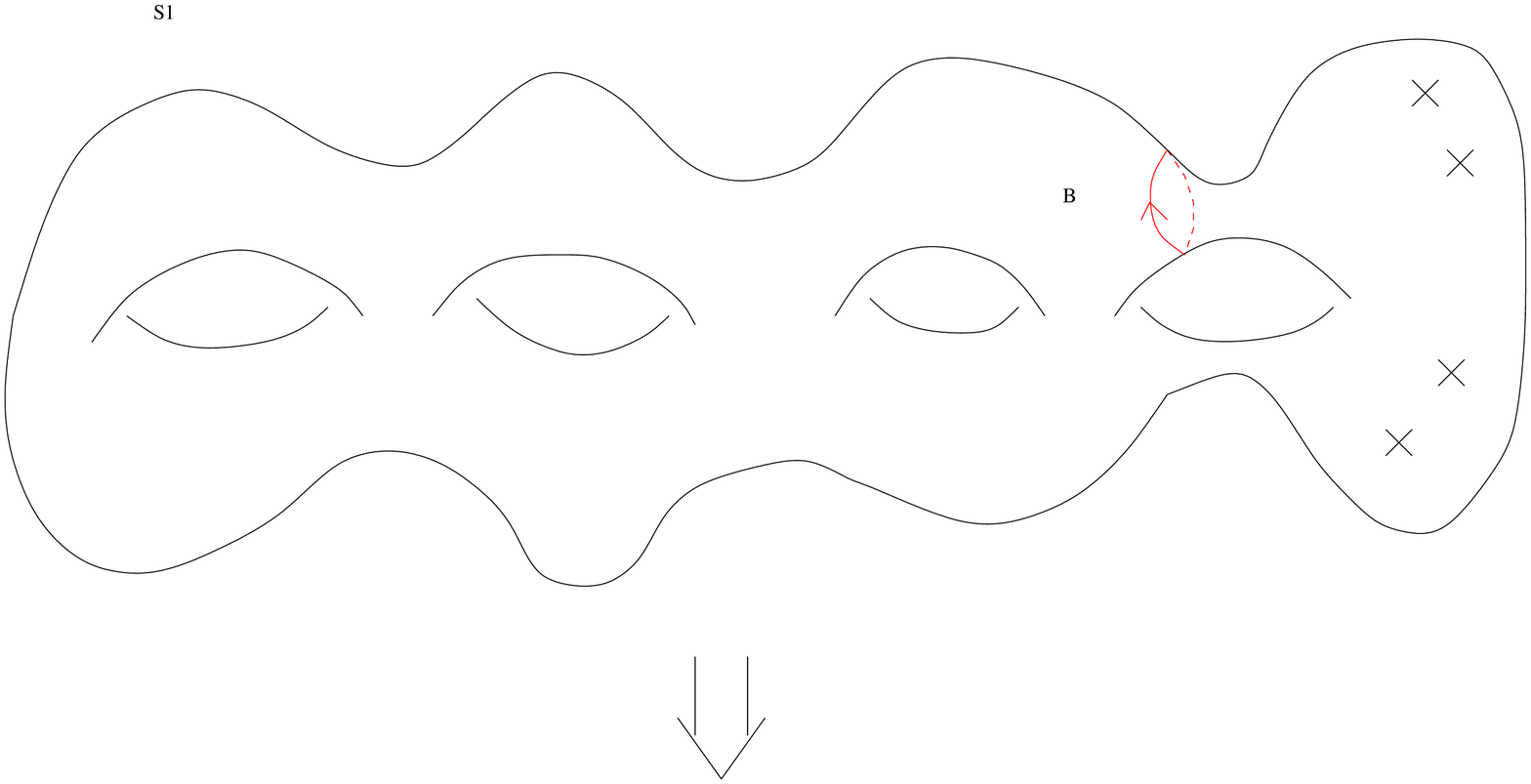}
\label{Sfig3b}
\centering
\psfrag{B}{\footnotesize{${\mathcal{C}}$}} 
\psfrag{S1}{\footnotesize{$\tilde{\Sigma}_{1}$}}
\psfrag{S2}{\footnotesize{$\tilde{\Sigma}_{2}$}}
\psfrag{CI}{\footnotesize{$C^{\pm}_{I}$}}
\psfrag{Ip}{\footnotesize{$\infty^{+}$}}
\psfrag{Im}{\footnotesize{$\infty^{-}$}}
\includegraphics[width=100mm]{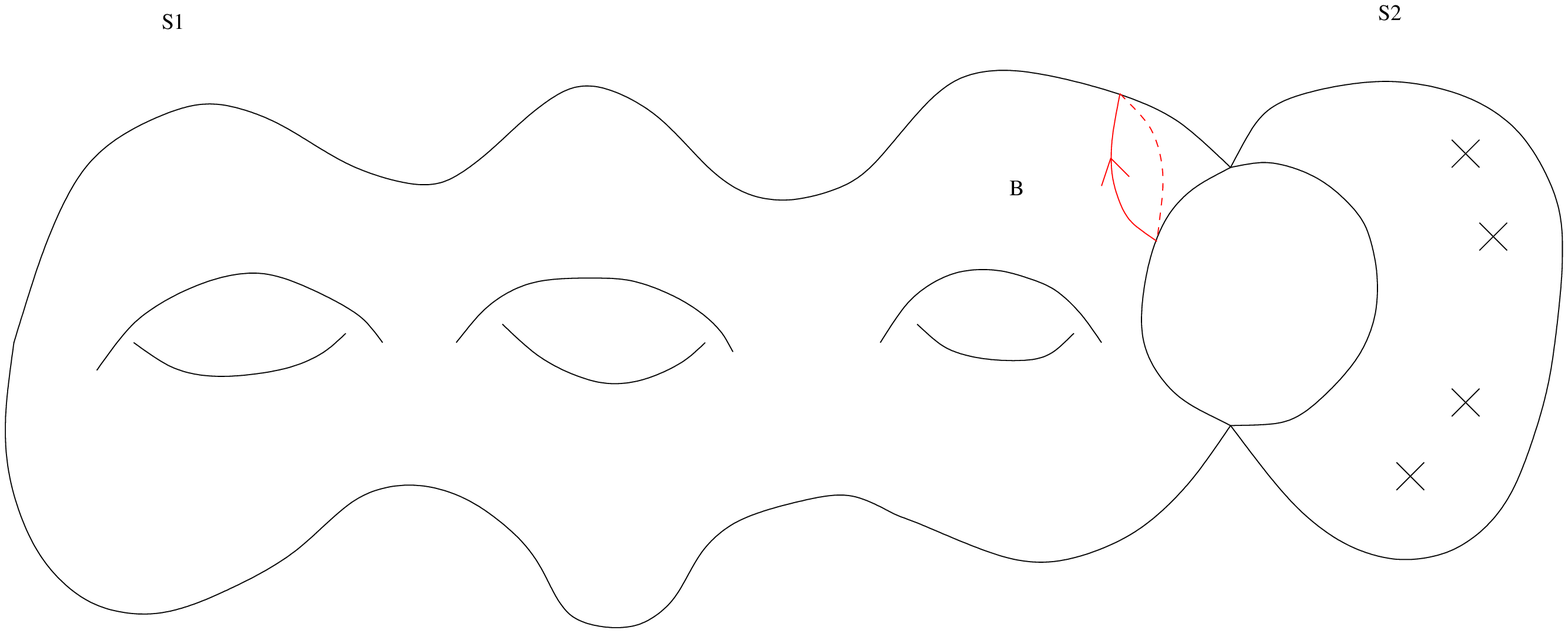}
\caption{The degeneration $\Sigma \rightarrow \tilde{\Sigma}_{1}\cup
\tilde{\Sigma}_{2}$. The four singular points $\pm 1^{\pm}$ are marked
with crosses on the curve}
\label{Sfig3}
\end{figure}
\paragraph{}
Starting from the original spectral curve,
\bea 
\Sigma\,\,:\qquad{} y^{2} & = & \left(x-b_{+}\right)  
\left(x-b_{-}\right) \prod_{i=1}^{K-1} 
\left(x-a^{(i)}_{+}\right)\left(x-a^{(i)}_{-}\right)
\nn \eea
the curve $\tilde{\Sigma}_{1}$ is obtained by ``blowing-up'' the
region near $x=\infty$. Thus we set, 
\bea x\,=\,\rho\tilde{x} & \qquad{} & y\,=\,\rho^{K}\tilde{x}\tilde{y}_{1}
\label{cov}
\eea 
holding $\tilde{x}$ and $\tilde{y}_{1}$ fixed as $\rho\rightarrow
\infty$. Thus we obtain the curve, 
\bea 
\tilde{\Sigma}_{1}\,\,:\qquad{} \tilde{y}^{2}_{1} & = & \prod_{i=1}^{K-1} 
\left(\tilde{x}-\tilde{a}^{(i)}_{+}\right)\left(\tilde{x}-
\tilde{a}^{(i)}_{-}\right)
\nn \eea
This is a generic hyper-elliptic curve of genus $K-2$. It can be
represented as a double cover of the complex $\tilde{x}$-plane with
$K-2$ cuts, $\tilde{C}^{\pm}_{I}$, with $I=1,2,\ldots, K/2-1$, and  
$\tilde{C}_{0}$ arranged as shown in Figure 5. We introduce a
corresponding set of one-cycles, 
$\tilde{\mathcal{A}}^{\pm}_{I}$, $\tilde{\mathcal{A}}_{0}$ 
which encircle the cuts 
$\tilde{C}^{\pm}_{I}$ and $\tilde{C}_{0}$ respectively as shown in
Figure 6. The conjugate cycles $\tilde{\mathcal{B}}^{\pm}_{I}$,
$\tilde{\mathcal{B}}_{0}$ run from the point at infinity on the top sheet to
the point at infinity on the lower sheet, passing through the
corresponding cut as also shown in this figure. The curve also has 
punctures at the two points $0^{\pm}$ above $\tilde{x}=0$, which
correspond to the shrinking cycle\footnote{More precisely the
  resulting meromorphic differential $d\tilde{p}_{1}$ on
  $\tilde{\Sigma}_{1}$ 
discussed below has poles at these points.}.   
\begin{figure}
\centering
\psfrag{x}{\footnotesize{$\tilde{x}$}}
\psfrag{Z}{\footnotesize{$0$}}
\psfrag{C0}{\footnotesize{$\tilde{C}_{0}$}}
\psfrag{C1m}{\footnotesize{$\tilde{C}_{1}^{-}$}}
\psfrag{C1p}{\footnotesize{$\tilde{C}_{1}^{+}$}}
\psfrag{C1k}{\footnotesize{$\tilde{C}_{K/2-1}^{-}$}}
\psfrag{C1j}{\footnotesize{$\tilde{C}_{K/2-1}^{+}$}}
\psfrag{a}{\footnotesize{$\tilde{a}^{(K-1)}_{-}$}}
\psfrag{b}{\footnotesize{$\tilde{a}^{(K-2)}_{-}$}}
\psfrag{g}{\footnotesize{$\tilde{a}^{(K-2)}_{+}$}}
\psfrag{h}{\footnotesize{$\tilde{a}^{(K-1)}_{+}$}}
\psfrag{k}{\footnotesize{$\tilde{a}^{(1)}_{-}$}}
\psfrag{l}{\footnotesize{$\tilde{a}^{(1)}_{+}$}}
\includegraphics[width=100mm]{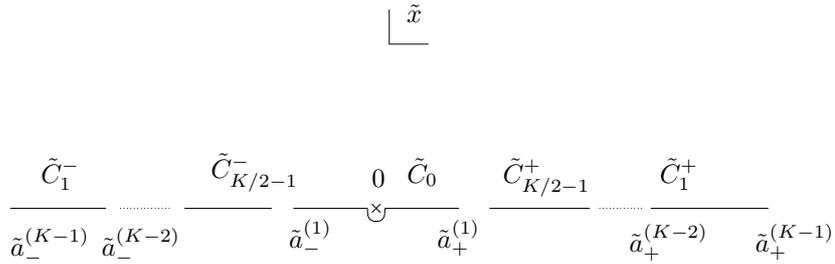}
\caption{The cut $\tilde{x}$-plane corresponding to the curve 
$\tilde{\Sigma}_{1}$}
\label{Sfig4}
\end{figure}
\begin{figure}
\centering
\psfrag{AI}{\footnotesize{$\tilde{\mathcal{A}}^{\pm}_{I}\, 
(\tilde{\mathcal{A}}_{0})$}}
\psfrag{BI}{\footnotesize{$\tilde{\mathcal{B}}^{\pm}_{I}\, 
(\tilde{\mathcal{B}}_{0})$}}
\psfrag{CI}{\footnotesize{$\tilde{C}^{\pm}_{I}\, 
(\tilde{C}_{0})$}}
\psfrag{Ip}{\footnotesize{$\infty^{+}$}}
\psfrag{Im}{\footnotesize{$\infty^{-}$}}
\includegraphics[width=100mm]{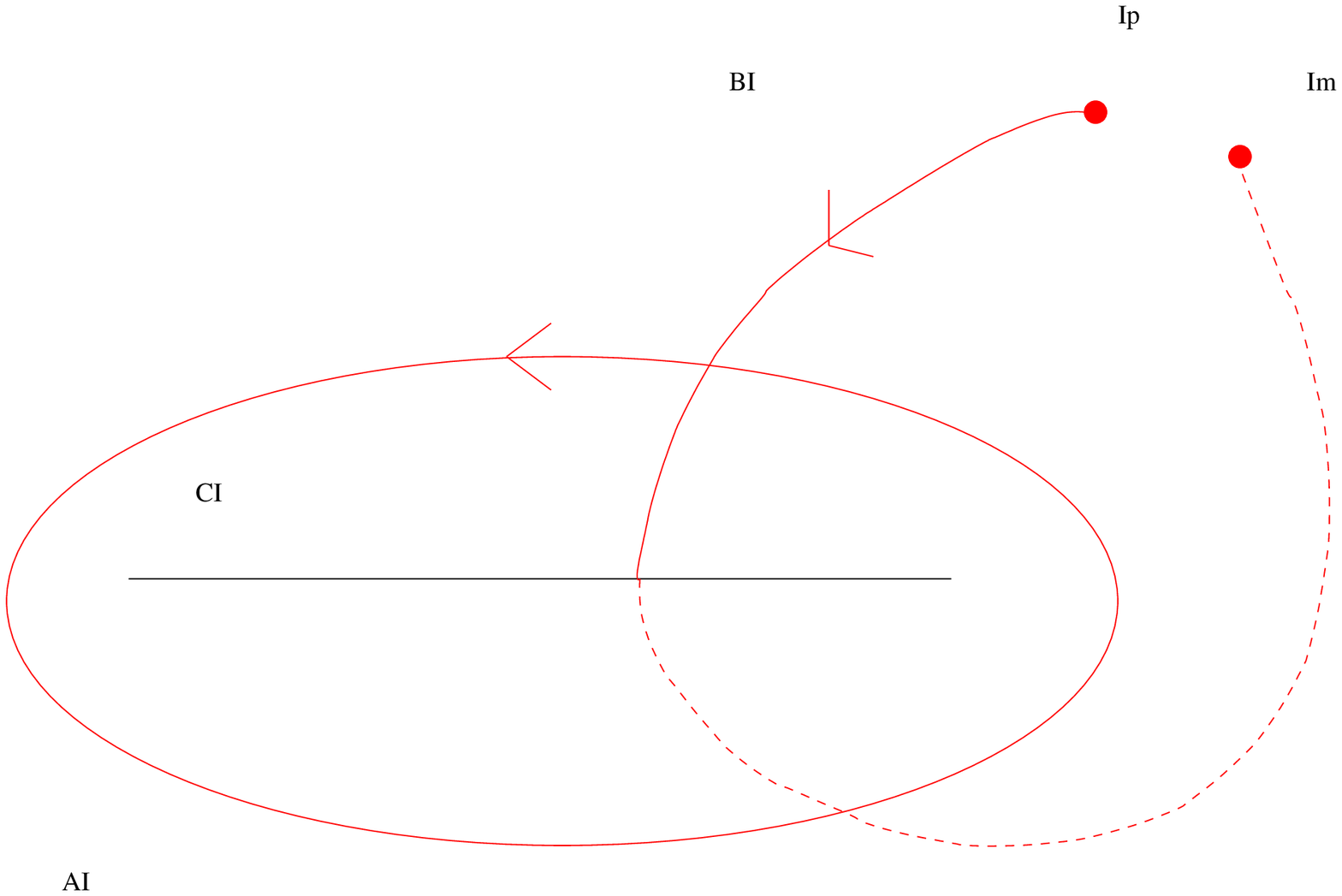}
\caption{The cycles on $\tilde{\Sigma}_{1}$. The
  index $I$ runs from $1$ to $K/2-1$.}
\label{Sfig5}
\end{figure}
\paragraph{}
We now consider the $\rho\rightarrow\infty$ limit of the differential
$dp=dp_{1}+dp_{2}$. The first term in (\ref{ansatz}), denoted
$dp_{1}$, involves $K-1$ arbitrary
constants $C_{\ell}$, $\ell=0,\ldots,K-2$. Its limiting behaviour is, 
\bea 
dp_{1} & \rightarrow & -\frac{d\tilde{x}}{\tilde{y}_{1}}\,
\sum_{\ell=0}^{K-2}   
\rho^{\ell+1-K}C_{\ell}\tilde{x}^{\ell-1} \nn \eea 
We will choose to scale the undetermined coefficient $C_{\ell}$ so as
to retain $K-1$ free parameters in the
resulting differential on $\tilde{\Sigma}_{1}$. Thus we set 
$C_{\ell}=\tilde{C}_{\ell}\rho^{K-\ell-1}$ and hold $\tilde{C}_{\ell}$
fixed. In this case $dp_{1}$ has a finite limit as $\rho\rightarrow
\infty$ while $dp_{2}\rightarrow 0$. The net result is, 
\bea
dp\,=\,dp_{1}+dp_{2} & \rightarrow & d\tilde{p}_{1}\,=\, 
-\frac{d\tilde{x}}{\tilde{y}_{1}}\,\sum_{\ell=0}^{K-2}   
\tilde{C}_{\ell}\tilde{x}^{\ell-1} \label{ansatz2} \eea
This is a meormorphic differential on $\tilde{\Sigma}_{1}$. It has simple
poles at the punctures ${0}^{\pm}$ above $\tilde{x}=0$ with residues 
$\pm \tilde{C}_{0}/\tilde{Q}$ and no other singularities on
$\tilde{\Sigma}_{1}$.  
\paragraph{}
It is easy to check that all but one of 
the periods of $dp$ on $\Sigma$ go over to corresponding periods of 
$d\tilde{p}_{1}$ on $\tilde{\Sigma}_{1}$. In particular we find, 
\bea 
\lim_{\rho\rightarrow\infty}\,\left[\oint_{\mathcal{A}^{\pm}_{I}}\,
  dp\,\right] \,\,=\,\, \oint_{\tilde{\mathcal{A}}^{\pm}_{I}}\,
  d\tilde{p}_{1} & \qquad{} & \lim_{\rho\rightarrow\infty}\,
\left[\oint_{\mathcal{B}^{\pm}_{I}}\,
  dp\,\right] \,\,=\,\, \oint_{\tilde{\mathcal{B}}^{\pm}_{I}}\,
  d\tilde{p}_{1} \nn \eea
for $I=1,2,\ldots,K/2-1$ and also, 
\bea 
\lim_{\rho\rightarrow\infty}\,\left[\oint_{\bar{\mathcal{A}}}\,
  dp\,\right] \,\,=\,\, \oint_{\tilde{\mathcal{A}}_{0}}\,
  d\tilde{p}_{1} & \qquad{} & \lim_{\rho\rightarrow\infty}\,
\left[\oint_{{\mathcal{B}^{\pm}_{K/2}}}\,
  dp\,\right] \,\,=\,\,\pm\, \oint_{\tilde{\mathcal{B}}_{0}}\,
  d\tilde{p}_{1} \nn \eea
where $\bar{\mathcal{A}}\,=\, 
\tilde{\mathcal{A}}^{+}_{K/2}+ \tilde{\mathcal{A}}^{-}_{K/2}$. The
above results are straightforwardly obtained by making the change of
variables (\ref{cov}) in each period integral and keeping only the
leading contribution as $\rho\rightarrow \infty$.  
\paragraph{}
As shown in Figure 4, the vanishing cycle 
$\hat{\mathcal{B}}={\mathcal{B}}^{+}_{K/2}- 
{\mathcal{B}}^{-}_{K/2}$ becomes a closed contour $\mathcal{C}$ 
surrounding the marked point on the top sheet above $\tilde{x}=0$, so
we also have, 
 \bea 
\lim_{\rho\rightarrow\infty}\,\left[\oint_{\bar{\mathcal{A}}}\,
  dp\,\right]  & = & \oint_{\mathcal{C}}\, d\tilde{p}_{1} = 2\pi K
\label{eqq} \eea 
Comparing with (\ref{ansatz2}), we see that this integral is equal to
the residue of $d\tilde{p}_{1}$ at the point ${0}^{+}$. Then Eqn 
(\ref{eqq}) is solved by setting $\tilde{C}_{0}=\tilde{Q}K$. 
\paragraph{} 
To summarise the above discussion the defining conditions for the
differential $dp$ on $\Sigma$ have reduced to a set of conditions for
the meromorphic differential $d\tilde{p}_{1}$ on $\tilde{\Sigma}_{1}$,  
\begin{itemize}
\item{} $d\tilde{p}_{1}$ has {\em simple poles} at the points 
${O}^{\pm}$ above $\tilde{x}=0$ with residues $\pm K/i$, and no
  other singularities on $\tilde{\Sigma}$. Thus, 
\bea 
d\tilde{p}_{1} & \longrightarrow & \pm \,\,\frac{K}{i}\,
\frac{d\tilde{x}}{\tilde{x}} \qquad{} {\rm as}
\,\,\,\,\tilde{x}\rightarrow 0 \nn \eea 
\item{} $d\tilde{p}_{1}$ obeys the normalisation conditions, 
\bea 
\oint_{\tilde{\mathcal{A}}^{\pm}_{I}} \,d\tilde{p}_{1} \,\,=\,\,0   & 
\qquad{} & 
\oint_{\tilde{\mathcal{B}}^{\pm}_{I}} \,d\tilde{p}_{1} \,\,=\,\,\pm 2\pi I  
\label{norm2} \eea 
for $I=1,2,\ldots,K/2-1$ and, 
\bea 
\oint_{\tilde{\mathcal{A}}^{0}} \,d\tilde{p}_{1}\,\,=\,\,0   & \qquad{} & 
\oint_{\tilde{\mathcal{B}}^{0}} \,d\tilde{p}_{1} \,\,=\,\, \pi \,K  
\label{norm3} \eea
\end{itemize}   
\paragraph{}
The second component in (\ref{degen}), the curve $\tilde{\Sigma}_{2}$, 
arises from blowing up the region around the points $0^{\pm}$
above $x=0$. We scale the coordinates as, 
\bea y\,=\, \tilde{Q}\rho^{K-1}\tilde{y}_{2} & \qquad & {\rm with}\,\,\,\, 
\tilde{Q}^{2}\,=\,\prod_{i=1}^{K-1}\,\tilde{a}^{(i)}_{+}\tilde{a}^{(i)}_{-} 
\label{Qtilde} \eea
and take the limit $\rho\rightarrow \infty$ with $x$ and
$\tilde{y}_{2}$ held fixed to get the curve,  
\bea 
\tilde{\Sigma}_{2}\,\,:\qquad{} \tilde{y}^{2}_{2} & = &
x^{2}\,-\,b^{2} 
\nn \eea
which has genus zero. This curve contains the original four singular points
$\pm 1^{\pm}$ and also has two new punctures at the points
$\infty^{\pm}$ corresponding to the vanishing cycles. 
\paragraph{}
The differential $dp_{2}$ on $\Sigma$ gives rise to the following
meromorphic differential on $\tilde{\Sigma}_{2}$ with double poles at
the points over $x=\pm 1$; 
\bea 
d\tilde{p}_{2} & = & -\,\frac{i\pi J}{\sqrt{\lambda}}\left[ 
\sqrt{b^{2}-1}\left(\frac{1}{(x-1)^{2}}\,+\,\frac{1}{(x+1)^{2}}\right)
\,\,-\,\, 
\frac{1}{\sqrt{b^{2}-1}}\left(\frac{1}{(x-1)}\,-\,\frac{1}{(x+1)}\right)
\,\right]\,\,\frac{dx}{\tilde{y}_{2}}\nn  \\ \label{dp2} \eea
\paragraph{}
\begin{figure}
\centering
\psfrag{S1}{\footnotesize{$\Sigma$}}
\psfrag{S2}{\footnotesize{$\tilde{\Sigma}_{2}$}}
\psfrag{Ah}{\footnotesize{$\mathcal{A}^{+}_{K/2}$}} 
\psfrag{A2}{\footnotesize{$\hat{\mathcal{A}}_{2}$}}
\psfrag{Ip}{\footnotesize{$\infty^{+}$}}
\psfrag{Im}{\footnotesize{$\infty^{-}$}}
\includegraphics[width=100mm]{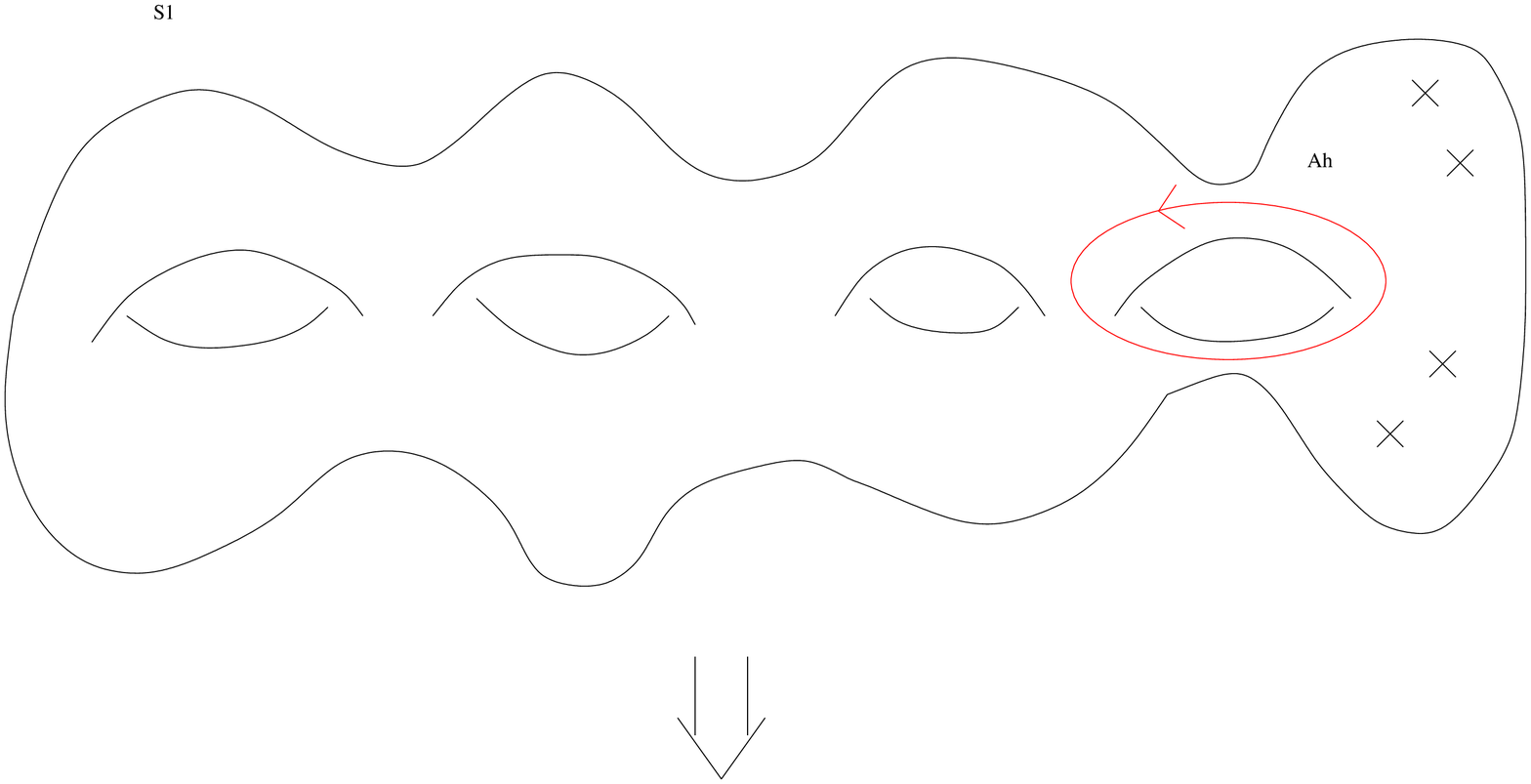}
\label{Sfig6b}
\centering
\psfrag{S1}{\footnotesize{$\tilde{\Sigma}_{1}$}}
\psfrag{S2}{\footnotesize{$\tilde{\Sigma}_{2}$}}
\psfrag{A1}{\footnotesize{$\hat{\mathcal{A}}_{1}$}} 
\psfrag{A2}{\footnotesize{$\hat{\mathcal{A}}_{2}$}}
\psfrag{Ip}{\footnotesize{$\infty^{+}$}}
\psfrag{Im}{\footnotesize{$\infty^{-}$}}
\includegraphics[width=100mm]{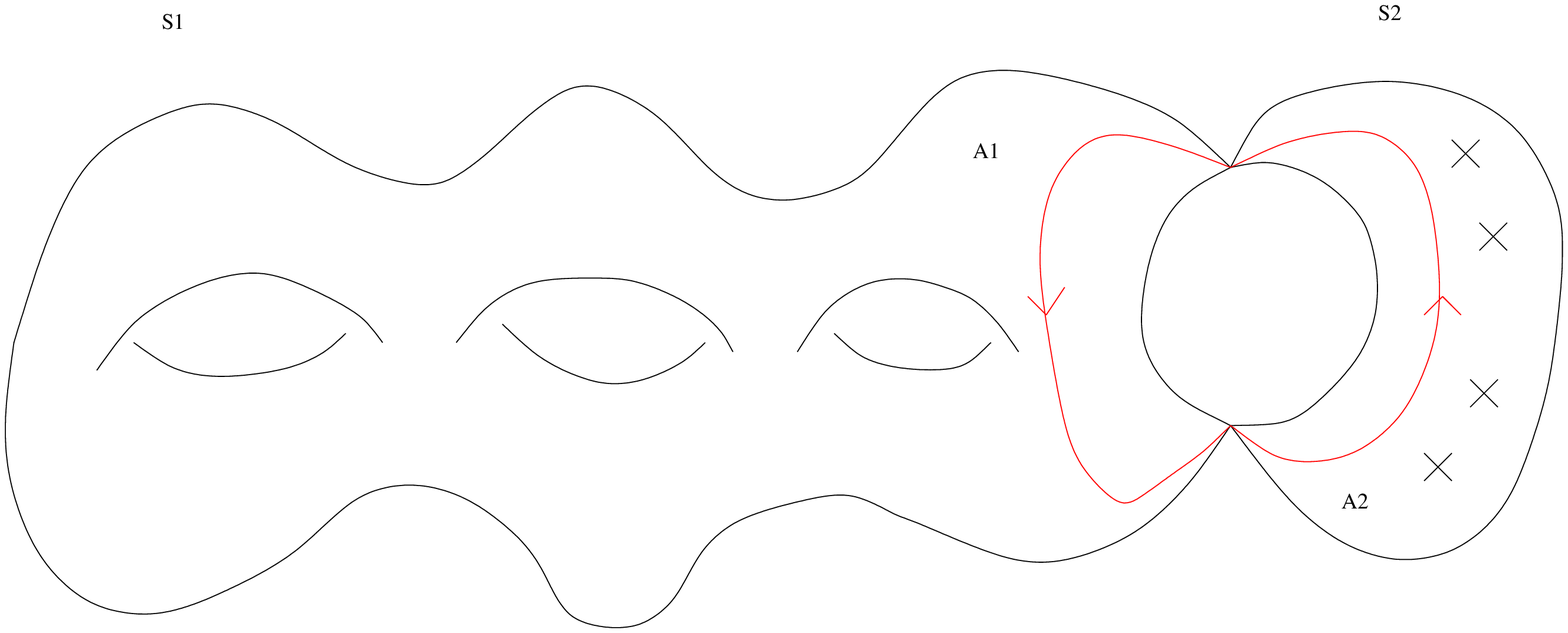}
\caption{The ``extra'' cycle $\mathcal{A}^{+}_{K/2}$ becomes the sum
  of the chains $\hat{\mathcal{A}}_{1}$ and $\hat{\mathcal{A}}_{2}$.}
\label{Sfig6}
\end{figure}
The surface $\tilde{\Sigma}_{2}$ enters when considering the limit of
the final period condition corresponding to the cycle
$\hat{\mathcal{A}}=\mathcal{A}_{K/2}^{+}$. 
The analysis of this condition is presented in the Appendix. In
particular we find the following limit for this equation, 
\bea
\lim_{\rho\rightarrow\infty}\,\left[\oint_{\hat{\mathcal{A}}}\,
  dp\,\right]  & = & 
\int_{\hat{\mathcal{A}}^{\rm reg}_{1}}\,d\tilde{p}_{1}\,+\, 
\int_{\hat{\mathcal{A}}_{2}}\,d\tilde{p}_{2}\,\,=\,\,0
\label{match} \eea
where $\hat{\mathcal{A}}^{\rm reg}_{1}$ and $\hat{\mathcal{A}}_{2}$ 
are suitably
regulated ``chains'' (ie open contours) on $\tilde{\Sigma}_{1}$ and
$\tilde{\Sigma}_{2}$ as shown in Figure 7. More precisely, 
we define,  
\bea 
\int_{\hat{\mathcal{A}}^{\rm reg}_{1}}\,d\tilde{p}_{1} \,\,= 
\,\, \int_{\epsilon^{-}}^{\epsilon^{+}} \,d\tilde{p}_{1} & \qquad{} & 
\int_{\hat{\mathcal{A}}_{2}}\,d\tilde{p}_{2}\,\,=
\,\, -\,\int_{\infty^{-}}^{\infty^{+}} \,\,
d\tilde{p}_{2} \nn \eea
where $\epsilon=b/\rho$ and ${\epsilon}^{\pm}$ are the two points
above $\tilde{x}=\epsilon$ on $\tilde{\Sigma}_{1}$ and $\infty^{\pm}$
are the two points above $x=\infty$ on $\tilde{\Sigma}_{2}$.  
\paragraph{}
Finally in the $\rho\rightarrow \infty$ limit, 
the conserved charges have the behaviour, 
\bea 
\Delta\,+\,S & \simeq & \frac{\sqrt{\lambda}}{2\pi} \,\,\rho 
\qquad{} \rightarrow 
\,\,\,\,\infty \nn \\ 
\Delta\,-\,S & \simeq & \frac{\sqrt{\lambda}}{2\pi} \,\, \frac{K}{b} 
\,\,\,+\,\,\,\frac{J}{b}\left(\sqrt{b^{2}-1}\,+\,
\frac{1}{\sqrt{b^{2}-1}}\right) \label{behave} \eea
up to corrections which vanish as $\rho\rightarrow \infty$. 
\section{The solution}
\paragraph{}
To solve for the spectrum in the large-$S$ limit we need determine the
pair $(\tilde{\Sigma}_{1}, d\tilde{p}_{1})$ and then solve the
matching condition (\ref{match}). The first task is similar in nature
to the original problem of finding $dp$, in that we must solve the
normalisation conditions for the meromorphic differential
$d\tilde{p}_{1}$ on a generic hyperelliptic curve
$\tilde{\Sigma}_{1}$. There is however, an important
difference: while the orginal differential $dp$ had double poles at
four points on $\Sigma$, the new differential $d\tilde{p}_{1}$ has
only simple poles with integral residues, 
\bea 
d\tilde{p}_{1} & \longrightarrow & \pm \,\,\frac{K}{i}\,
\frac{d\tilde{x}}{\tilde{x}} \qquad{} {\rm as}
\,\,\,\,\tilde{x}\rightarrow 0 \label{simp2} \eea 
and no other singularities. The resulting problem of reconstructing  
$(\tilde{\Sigma}_{1}, d\tilde{p}_{1})$ is then a standard one which
arises for example in the study of the F-terms of ${\cal
  N}=2$ SUSY gauge theories \cite{Vafa}\footnote{See in particular
  subsection 3.3 of this reference.}. We now describe its solution.   
\paragraph{}
Integrating (\ref{simp2}) we find that, 
\bea 
\tilde{p}_{1}(\tilde{x}) & \longrightarrow & \pm \,\,\frac{K}{i}\,
\log\tilde{x} \qquad{} {\rm as}
\,\,\,\,\tilde{x}\rightarrow 0 \nn \eea 
and thus we have, 
\bea 
\exp\left(\pm i\tilde{p}_{1}(\tilde{x})\right) & \longrightarrow & 
\left(\tilde{x}\right)^{\pm K}\, \qquad{} {\rm as}
\,\,\,\,\tilde{x}\rightarrow 0 \label{pole1} \eea 
\paragraph{}
Now consider the function, 
\bea 
f(\tilde{x}) & = & 2\cos\tilde{p}_{1}(\tilde{x}) \,\,=\,\,
\exp\left(+i\tilde{p}_{1}(\tilde{x})\right)\,+\,
\exp\left(-i\tilde{p}_{1}(\tilde{x})\right)
\nn \eea
As the periods of $d\tilde{p}_{1}$ are normalised in integer units 
$f$ is analytic on the the complex $\tilde{x}$ plane. According to 
Eqn (\ref{pole1}) it has a pole of order $K$ at $\tilde{x}=0$ and no
other singularities. Its behaviour at infinity is inherited from that
of $p(x)$; 
\bea \tilde{p}_{1}(\tilde{x}) \rightarrow 0 &\,\, & {\rm as} \,\,\,\,\tilde{x}
\rightarrow \infty \nn \eea 
and thus $f\rightarrow 2$ as $\tilde{x}\rightarrow \infty$. The most
general analytic function obeying these conditions can be
parameterised in terms of $K-1$ undetermined coefficients
$\tilde{q}_{j}$, with $j=2,\ldots,K$, as,  
\bea f(\tilde{x})\,=\, \mathbb{P}\left(\frac{1}{\tilde{x}}\right) & =
& 2\,\,+\,\,\frac{\tilde{q}_{2}}
{\tilde{x}^{2}}\,\,+\,\, \frac{\tilde{q}_{3}}
{\tilde{x}^{3}}\,\,+\,\,\ldots\,\,+\,\, \frac{\tilde{q}_{K}}
{\tilde{x}^{K}} \label{f} \eea
Thus we have an explicit solution for 
$\tilde{p}_{1}(\tilde{x})=\cos^{-1}(f/2)$ which yields a meromorphic
differential, 
\bea d\tilde{p}_{1} & = & -i\,\frac{d\tilde{x}}{\tilde{x}^{2}}\,\,
\frac{{\mathbb{P}}'_{K}\left(\frac{1}{\tilde{x}}\right)}
{ \sqrt{{\mathbb{P}}_{K}
\left(\frac{1}{\tilde{x}}\right)^{2}\,\,-\,\,4}} 
\label{dp1} \eea 
One may easily check that this differential satisfies the
normalisation conditions (\ref{norm2},\ref{norm3}) and has a poles 
with the required residue at the points over $\tilde{x}=0$.  
The differential $d\tilde{p}_{1}$ is meromorphic on the curve, 
\bea 
\tilde{\Sigma}_{1}\,\,:\qquad{} \tilde{y}^{2}_{1} & = & \prod_{i=1}^{K-1} 
\left(\tilde{x}-\tilde{a}^{(i)}_{+}\right)\left(\tilde{x}-
\tilde{a}^{(i)}_{-}\right)
\nn \\ 
& = & \frac{\tilde{x}^{2K}}{4\tilde{q}_{2}}\left[{\mathbb{P}}_{K}
\left(\frac{1}{\tilde{x}}\right)^{2}\,\,-\,\,4\right] \nn  \eea 
and we can rewrite (\ref{dp1}) as, 
\bea 
d\tilde{p}_{1}\,=\, -\frac{d\tilde{x}}{\tilde{y}_{1}}\,
\sum_{\ell=0}^{K-2}\,\tilde{C}_{\ell} \, \tilde{x}^{\ell-1} & {\rm
  with} & \tilde{C}_{\ell} \, =\, -\frac{(K-\ell)\tilde{q}_{K-\ell}}{2
\sqrt{-\tilde{q}_{2}}} \nn \eea 
Thus we have expressed the $2K-2$ parameters corresponding to the 
branch-points $\tilde{a}^{(i)}_{\pm}$ of the curve
$\tilde{\Sigma}_{1}$ and the $K-1$ parameters
corresponding to the undetermined coefficients $\tilde{C}_{\ell}$ in
the differential $d\tilde{p}_{1}$ in terms $K-1$
parameters $\tilde{q}_{j}$, $j=2,\ldots,K$. At this point we observe
that the curve $\tilde{\Sigma}_{1}$ and differential $d\tilde{p}_{1}$ 
are essentially identical to the curve $\Gamma_{K}$ and differential
$d\hat{p}$ of the $SL(2,\mathbb{R})$ spin chain. 
\paragraph{}
The matching condition,  
\bea 
\tilde{p}_{1}(\tilde{x}=\epsilon) & = & -\frac{1}{2}
\int_{\infty^{-}}^{\infty^{+}}\,d\tilde{p}_{2} \nn \eea
with $\epsilon=b/\rho$ can now be evaluated explicitly in terms of the
closed formulae (\ref{dp1},\ref{dp2}) for the differentials
$d\tilde{p}_{1}$ and $d\tilde{p}_{2}$. It yields, 
\bea \frac{1}{i} \log\left(\frac{ \rho^{K} \tilde{q}_{K}}{b}\right) & = &  
\frac{2\pi i J}{\sqrt{\lambda}} \frac{1}{\sqrt{b^{2}-1}} \nn \eea 
or equivalently, 
\bea \sqrt{b^{2}-1} & = & \frac{2\pi J}{\sqrt{\lambda}}\,\times \, 
\frac{1}{  K\log \left(\rho \tilde{q}_{K}^{1/K}\right)} \label{match2}
\eea 
Thus $b\rightarrow 1$ and 
the inner branch points approach the punctures at the points
$x=\pm 1$ as the scaling parameter $\rho$ goes to infinity. 
\paragraph{}
Finally we can evaluate the conserved charges in the limit
$\rho\rightarrow \infty$, 
\bea 
\Delta\,+\,S & = & \frac{\sqrt{\lambda}}{2\pi} \,\sqrt{-\tilde{q}_{2}}
\rho \qquad{} \rightarrow 
\,\,\,\,\infty \nn \\ 
\Delta\,-\,S & = & \frac{\sqrt{\lambda}}{2\pi} \,K 
\,\,\,+\,\,\,\frac{J}{\sqrt{b^{2}-1}} \label{dms2} \eea 
These relations confirm 
that $S\rightarrow\infty$ as $\rho\rightarrow\infty$ as anticipated. 
Using (\ref{match2}) we obtain,  
\bea \rho  & \simeq & \frac{4\pi}{\sqrt{\lambda}}\, \frac{S}
{\sqrt{-\tilde{q}_{2}}} \label{rho} \eea 
Eliminating $\rho$ from (\ref{dms2}) then gives, 
\bea 
\Delta\,-\,S & = & \frac{\sqrt{\lambda}}{2\pi}\,\left[K\log S
  \,\,+\,\, 
\log(\tilde{q}_{K}/\sqrt{-\tilde{q}_{2}})\right] \,\,+\,\,O(1/\log S)
\label{dms} \eea
\paragraph{}
In classical string theory the parameters $\tilde{q}_{i}$ are
continuous variables. 
To complete the solution of the model we must
also consider the semiclassical quantisation conditions \cite{DV1,
  DV2}. As mentioned
above, semiclassical
quantization of string theory on $AdS_{3}\times S^{1}$ is accomplished
by quantizing the filling fractions in integer units,  
\bea 
\mathcal{S}^{\pm}_{I} & = & -\frac{1}{2\pi i}\,\cdot\,
\frac{\sqrt{\lambda}}{4\pi}\,\,\oint_{\mathcal{A}^{\pm}_{I}}\,
\left(x\,+\,\frac{1}{x}\right)\,dp\,\,=\,\, 
l^{\pm}_{I}\in \,\,\mathbb{Z}^{+} \label{quant}
\eea
for $I=1,2,\ldots,K/2$. The integers $l^{\pm}_{I}$ obey, 
\bea 
\sum_{I=1}^{K/2} \, \left(l^{+}_{I}\,+\,l^{-}_{I}\right)\,\,=\,\,S & 
\qquad{} & \sum_{I=1}^{K/2} \,I\left( l^{+}_{I}\,-\,l^{-}_{I}\right)\,\,=\,\,0
\label{lcond} 
\eea
\paragraph{}
We now consider the limiting form of these
conditions in the scaling limit $\rho\rightarrow \infty$. This is
easily implimented by setting $x=\rho\tilde{x}$ in the integrals
appearing on the LHS of (\ref{quant}) holding $\tilde{x}$ fixed in the
limit. In this case the periods of the differential $(x+1/x)dp$ on
$\Sigma$ go over to periods of $\tilde{x}d\tilde{p}_{1}$ on 
$\tilde{\Sigma}_{1}$ as $\rho\rightarrow \infty$. In particular we
find, 
\bea 
 -\frac{1}{2\pi i}\,\cdot\,
\frac{1}{\sqrt{-\tilde{q}_{2}}}\,\,\oint_{\tilde{\mathcal{A}}^{\pm}_{I}}\,
\tilde{x}\,d\tilde{p}_{1}  & = & \frac{l^{\pm}_{I}}{S} \label{q1}
\eea 
for $I=1,2,\ldots,K/2-1$ and, 
\bea 
 -\frac{1}{2\pi i}\,\cdot\,
\frac{1}{\sqrt{-\tilde{q}_{2}}}\,\,\oint_{\tilde{\mathcal{A}}_{0}}\,
\tilde{x}\,d\tilde{p}_{1}  & = & \frac{\bar{l}}{S} \label{q2}
\eea 
with $\bar{l}=l^{+}_{K/2}+l^{-}_{K/2}$. 
\paragraph{}
The quantization conditions (\ref{q1},\ref{q2}) lead to a discrete
spectrum labeled by the $K-1$ integers $l^{\pm}_{I}$ and $\bar{l}$, 
\bea 
\gamma
\left[l_{1}^{+},l_{1}^{-},\ldots, l_{K/2-1}^{+},l_{K/2-1}^{-},
    \bar{l}\right] & \simeq & \frac{\sqrt{\lambda}}{2\pi}\left[ K\log\,S
    \,\,+\,\, \log\left(\frac{\tilde{q}_{K}}{\left(-\tilde{q}_{2}
\right)^{\frac{K}{2}}}\right)
    \,\,+\,\,C\,\,+\,\,O\left(\frac{1}{\log\,S}\right) \right] \nn \\ 
\label{fnal}
  \eea 
Finally one may check that the spectrum defined by equations
(\ref{q1},\ref{q2},\ref{fnal}) is identical to the result 
(\ref{stringspec}) given in the Introduction 
with the identifications, 
\bea \hat{q}_{j} & = & \frac{\tilde{q}_{j}}{\left(-\tilde{q}_{2}
\right)^{\frac{K}{2}}} \nn \eea 
for $j=2,\ldots,K$, $H_{K}=\log \hat{q}_{K}$ and, 
\bea 
l_{j} & = & l^{+}_{j} \qquad{} j=1,\ldots ,K/2 -1 \nn \\ 
& = & \bar{l} \qquad{} j=K/2 \nn \\ 
& = & l^{-}_{K-j} \qquad{} j=K/2+1,\ldots, K-1 \nn \eea 
In particular the conditions (\ref{lcond}) ensure that the integers 
$l_{j}$ obey the corresponding relations (\ref{BS1}). 
Thus we see that the spectra of one-loop gauge
theory and string theory differ only in the overall $\lambda$
dependent prefactor which takes the value $\sqrt{\lambda}/2\pi$ in
semiclassical string theory and $\lambda/4\pi^{2}$ in one-loop gauge theory. 
\paragraph{}
Finally, one feature of the gauge theory results which remains unclear on
the string side is the bound $K\leq J$. In fact the large-$S$ string spectrum
derived above does not depend on $J$ at all. 
Despite this, our semiclassical analysis formally requires $J\sim
\sqrt{\lambda}>>1$. Thus the upper bound is therefore 
reached for solutions with
$K$ spikes only when $K\sim
\sqrt{\lambda}>>1$. It is unclear whether higher-loop worldsheet
corrections remain supressed when $K$ scales with $\lambda$ in this
way and it may therefore require a more sophisticated analysis
than the one presented above to detect the presence of an upper bound
on $K$ in string theory.       

\section{Interpretation}
\begin{figure}
\centering
\psfrag{AI}{\footnotesize{$\mathcal{A}^{\pm}_{I}$}}
\psfrag{BI}{\footnotesize{$\mathcal{B}^{\pm}_{I}$}}
\psfrag{CI}{\footnotesize{$C^{\pm}_{I}$}}
\psfrag{Ip}{\footnotesize{$\infty^{+}$}}
\psfrag{Im}{\footnotesize{$\infty^{-}$}}
\includegraphics[width=100mm]{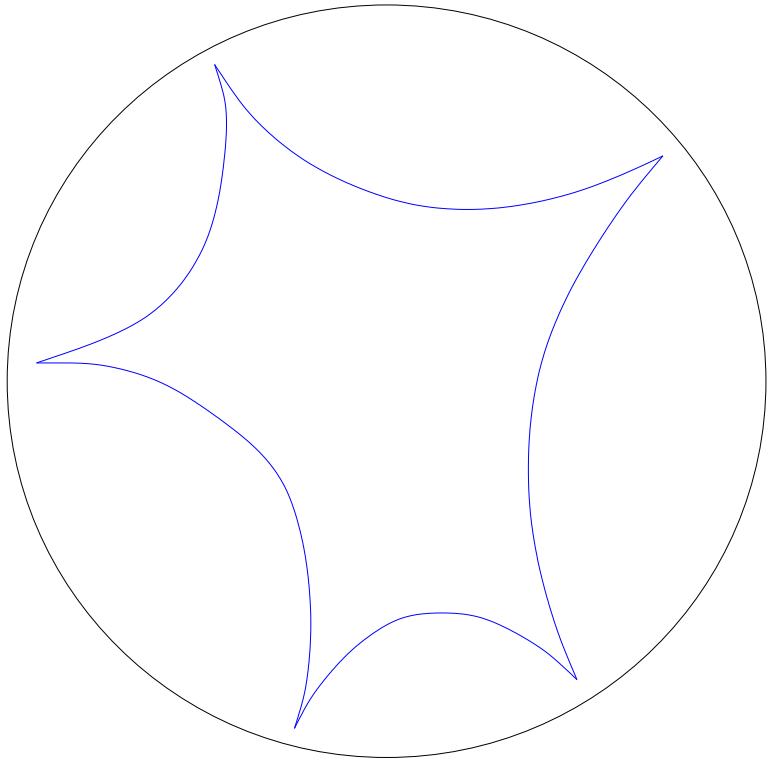}
\caption{A spinning string in $AdS_{3}$ with spikes approaching the boundary}
\label{Sfig7}
\end{figure}
\paragraph{}
In the minimal case $K=2$, the finite gap solution (with symmetric
cuts) considered above
reduces to the folded GKP string. Logarithmic scaling in $S$ 
with prefactor $2\,\cdot\,\sqrt{\lambda}/2\pi$ 
arises when the two folding points of the
string approach the boundary of $AdS_{3}$. Following \cite{Kruc}, it is
natural to expect that $\log S$ scaling with prefactor 
$K\,\cdot\, \sqrt{\lambda}/2\pi$
corresponds to strings with $K$ spikes\footnote{In the $S\rightarrow
  \infty$ limit with fixed $J$, the motion on $S^{1}$ can be neglected
  so that the string effectively move in the two spatial dimensions
  of $AdS_{3}$ as do the solutions of \cite{Kruc}. More generally, 
  motion in the extra $S^{1}$ will tend to smooth out the spikes.} 
which approach the boundary as
$S\rightarrow \infty$. Solutions with $\mathbb{Z}_{K}$ symmetry,
where the spikes lie at the vertices of a regular polygon were
constructed in \cite{Kruc}. Based on the scaling limit of the finite
gap construction considered 
above, we expect to find ($K-2$)-parameter families of spikey solutions 
in this limit. 
More precisely there should be $K-2$ parameters
  corresponding to the conserved action variables for 
solutions with fixed $S$ and an
  additional $K-2$ corresponding to the initial values of the 
conjugate angle variables. Solutions are also labelled by 
an orientation angle $\psi_{0}$ canonically conjugate to $S$. 
The generic solution need not 
have the symmetric form
considered in \cite{Kruc}, but rather should have variable angular
seperations between the spikes as shown in Figure 8. 
The details of this picture will be
presented elsewhere and will not be needed for the following
arguments.         
\paragraph{}
To analyse the $S\rightarrow\infty$ 
limit it will be convenient to use the representation
of $AdS_{3}$ as the group manifold $SU(1,1)$ with complex coordinates, 
$Z_{1}$, $Z_{2}$ obeying $|Z_{1}|^{2}-|Z_{2}|^{2}=1$. The complex
coordinates are related to the standard
global coordinates $(t,\rho,\psi)$ on $AdS_{3}$ by,  
\bea
Z_{1}\,=\, \cosh\rho\,\exp(it) & \qquad{} & Z_{2}\,=\,
\cosh\rho\,\exp(i\psi) \nn \eea
\paragraph{}
We introduce the group-valued worldsheet field, 
\bea 
g(\sigma,\tau) & = & \left(\begin{array}{cc} Z_{1} & Z_{2} \\ 
\bar{Z}_{2} & \bar{Z}_{1} \end{array} \right) \,\,\in\,\,SU(1,1) \nn
\eea 
and the conserved current corresponding to right multiplication in the
group,   
\bea j_{\pm}(\sigma,\tau)\,\,=\,\,g^{-1}\,\partial_{\pm}g & = &
\frac{1}{2}\eta_{AB} j^{A}s^{B}
\nn \eea
Here $s^{A}$, with $A=0,1,2$ are generators chosen to satisfy, 
 satisfying, 
\bea [s^{A},\,s^{B}] & = & -2\varepsilon^{ABC}s_{C} \nn \eea 
and, 
\bea \eta^{AB} & = & \frac{1}{2}{\rm tr}_{2}[s^{A}s^{B}] \nn \eea 
where $\eta={\rm diag}(-1,1,1)$ is the Killing form of the Lie algebra 
$su(1,1)$ which is used to raise and lower indices with the usual
summation convention. For a generic Lie algebra valued quantity, 
\bea 
X & = & \frac{1}{2}\eta_{AB} X^{A}s^{B} \nn \eea 
we will sometimes use vector notation $\vec{X}=(X^{0},X^{1},X^{2})$ 
In the following we will use the explict choice
$(s^{0},s^{1},s^{2})=(-i\sigma_{3},
\sigma_{1},-\sigma_{2})$ where $\sigma_{i}$ are the usual Pauli
matrices. 
\paragraph{}
The Noether charge corresponding to right multiplication is, 
\bea Q_{R}\,\,=\,\,\frac{1}{2}\eta_{AB}Q^{A}_{R}s^{B} & = & 
\frac{\sqrt{\lambda}}{4\pi}\,\int_{0}^{2\pi}\,d\sigma
\,j_{\tau}\,\,\,\,\in \,\,\,su(1,1) \nn \eea 
The Cartan generator is, 
\bea Q^{0}_{R}\,\,=\,\, \Delta\,+\,S & = & 
\frac{\sqrt{\lambda}}{4\pi}\,\int_{0}^{2\pi}\,d\sigma
\,j^{0}_{\tau} \label{norm5} \eea 
We will focus on states of highest weight for which
$Q_{R}^{1}=Q_{R}^{2}=0$.
\paragraph{}
At fixed worldsheet time, we 
will assume that our solution has $K$ spikes at the points 
$\sigma=\sigma_{j}\in [0,2\pi]$ with $j=1,\ldots,K$. At these
points the $\sigma$-derivatives of all world-sheet fields vanish and
thus, 
\bea
j_{\pm} (\sigma=\sigma_{j},\tau)  & = & 
j_{\tau}(\sigma=\sigma_{j},\tau) \nn \eea
for all $j$. 
To understand the behaviour of the charge density near the spikes, we
consider the simplest two spike solution: the GKP folded string
\cite{GKP}. This
describes a folded string rotating around its midpoint in $AdS_{3}$. 
In global 
coordinates it has the form $t=\tau$, $\psi=\psi_{0}+\omega\tau$ (with
$\omega\geq 1$) and 
$\rho=\rho(\sigma)={\rm am}[i\tilde{\sigma}|\sqrt{1-\omega^{2}}]$ where, 
\bea \tilde{\sigma}= \frac{L}{2\pi}\,\sigma  & \qquad{} & L=
\frac{4}{\omega}\,\mathbb{K}\left(\frac{1}{\omega}\right) \nn \eea 
This is a two-parameter family of solutions labelled by $\omega$
(which determines $S$) and
$\psi_{0}$. 
The spikes are located at the points $\sigma=\sigma_{1}=\pi/2$ and 
$\sigma=\sigma_{2}=3\pi/2$. One may obtain the following
explicit form for the conserved current, 
\bea 
j^{0}_{\tau}(\sigma,\tau) & = & 
\frac{ 2\left[1+\frac{1}{\omega}{\rm sn}^{2}
\left(\frac{2\mathbb{K}}{\pi}\sigma\right. \left|\frac{1}{\omega}
\right)\right]}{{\rm dn}^{2}
\left(\frac{2\mathbb{K}}{\pi}\sigma\right. \left|\frac{1}{\omega}
\right)}   \nn \\ 
j^{1}_{\tau}(\sigma,\tau)\,+\,i\, j^{2}_{\tau}(\sigma,\tau) & = & 
-2\frac{\omega+1}{\omega}\,\exp\left(i\psi_{0}+i(\omega-1)\tau\right)\,\, 
\frac{{\rm sn}
\left(\frac{2\mathbb{K}}{\pi}\sigma\right. \left|\frac{1}{\omega}
\right)}{{\rm dn}^{2}
\left(\frac{2\mathbb{K}}{\pi}\sigma\right. \left|\frac{1}{\omega}
\right)}
\label{explicit}
\eea 
where $\mathbb{K}=\mathbb{K}(1/\omega)$. Conventions for elliptic
integrals and functions are as in \cite{BF}. 
\paragraph{}
The two spikes approach the boundary in the limit $\omega \rightarrow
1$. In this limit the conserved charges of the solution scale as,    
\bea 
S\simeq \,\frac{\sqrt{\lambda}}{2}\frac{\kappa}{\omega-1} \,\,+ \,\,\ldots &
  \qquad{} & \Delta-S \simeq\, \frac{\sqrt{\lambda}}{\pi} \log
\frac{1}{\omega-1}\,\,+\,\,\ldots \,\,\simeq  
\frac{\sqrt{\lambda}}{\pi} \log S\,\,+\,\,\ldots  \nn \eea 
and we define $\kappa^{-1}=\log(1/\sqrt{\omega-1})$. Thus the limit
$\omega\rightarrow 1$ (or equivalently, $\kappa\rightarrow 0$) implies 
$S\rightarrow \infty$. By inspection the current
$j_{\tau}^{0}(\sigma,\tau)$, 
which is the density of the conserved charge $\Delta+S\simeq 2S$, diverges as
$S\rightarrow \infty$. We define a normalised charge density, 
\bea 
\mu^{A}(\sigma,\tau) & = & \lim_{S\rightarrow \infty} \left[
  \frac{\sqrt{\lambda}}{8\pi S}\,j^{A}_{\tau}(\sigma,\tau)\right] \nn
\eea 
which remains finite and obeys, 
\bea 
\int_{0}^{2\pi} \, d\sigma \,\vec{\mu}(\sigma,\tau) & = &
\left(\begin{array}{c} 1 \\ 0 \\  0 \end{array}\right) \nn \eea 
for highest-weight states. 
\paragraph{}
Expanding around the spike point $\sigma_{1}$ we set, 
$\sigma=\sigma_{1}+\hat{\sigma}$ with $\hat{\sigma}<<1$ we find, 
\bea 
\mu^{0}(\sigma,\tau) & \simeq & \lim_{\kappa\rightarrow 0}\left[\frac{1}
{2\pi\kappa}\,
{\rm sech}^{2}\left(\frac{2\hat{\sigma}\kappa}{\pi}\right)\right]  \nn \\ 
& = & \frac{1}{2}\,\delta(\hat{\sigma}) \label{mu1} \eea 
and
\bea 
\mu^{1}(\sigma,\tau)\,+\, i\,\mu^{2}(\sigma,\tau) & \simeq & 
\lim_{\kappa\rightarrow 0}\left[-\frac{e^{i\psi_{0}}}{2\pi\kappa}\,
{\rm sech}^{2}\left(\frac{2\hat{\sigma}\kappa}{\pi}\right)\right]  \nn \\ 
& = & -\frac{e^{i\psi_{0}}}{2}\,\delta(\hat{\sigma}) \label{mu2} 
\eea 
\paragraph{}
The $\kappa\rightarrow 0$ limit leading to  (\ref{mu1},\ref{mu2}) 
focuses on the region of the string near the first spike. The
full charge density is obtained by including a similar contribution from the
second spike at $\sigma=\sigma_{2}$, 
\bea
\mu^{0}(\sigma,\tau)  & = &
\frac{1}{2}\delta(\sigma-\sigma_{1})\,\,+\,\,\frac{1}{2}
\delta(\sigma-\sigma_{2}) \nn  \\
\mu^{1}(\sigma,\tau)\,\,+\,\,i\,\mu^{2}(\sigma,\tau) 
& = & -\frac{e^{i\psi_{0}}}{2}\delta(\sigma-\sigma_{1})\,\,+\,\,\frac{
e^{i\psi_{0}}}{2} \delta(\sigma-\sigma_{2}) \nn \eea 
Equivalently we can write the large-$S$ limit of the current as, 
\bea \lim_{S \rightarrow\infty}\,
\left[{j}^{A}_{\tau}(\sigma,\tau)\right] & = &
\frac{8\pi}{\sqrt{\lambda}}\, \sum_{k=1}^{2}\,\,
L^{A}_{k}\,\delta\left(\sigma-\sigma_{k}\right)
\label{comp0} \eea
with 
\bea 
\vec{L}_{1}\,\,=\,\,\frac{S}{2}\left(\begin{array}{c} 1 \\
-\cos\psi_{0} \\ -\sin\psi_{0} \end{array}\right) & \qquad{} & 
\vec{L}_{2}\,\,=\,\,\frac{S}{2}\left(\begin{array}{c} 1 \\
\cos\psi_{0} \\ \sin\psi_{0} \end{array}\right) \label{l12} \eea
One can easily check that the highest weight 
conditions $Q_{R}^{1}=Q_{R}^{2}=0$ and the normalisation condition
(\ref{norm5}) are satisfied.  
\paragraph{}
The key feature of the above result is that the charge density becomes 
$\delta$-function localised at the spikes in the limit they approach
the boundary. We do not have much explicit information about solutions
for $K>2$ except in the $\mathbb{Z}_{K}$ symmetric case considered in 
\cite{Kruc}. Recently the symmetric solution of \cite{Kruc} has
  been analysed \cite{JJ} (see also \cite{JV}) 
in the same conformal gauge as we have just used to
  describe the GKP string. The behaviour in the vicinity of each spike
  is similar to that at the folds of the GKP string and in particular
  $\delta$-function localisation of the charge density will occur at
  each spike as it approaches the boundary. 
We will assume that the same is true for generic
  solutions with  $K$ spikes and thus we propose 
the obvious generalisation of (\ref{comp0}), 
\bea \lim_{S \rightarrow\infty}\,
\left[{j}^{A}_{\tau}(\sigma,\tau)\right] & = &
\frac{8\pi}{\sqrt{\lambda}}\, \sum_{k=1}^{K}\,\,
L^{A}_{k}\,\delta\left(\sigma-\sigma_{j}\right)
\label{comp} \eea 
where $L^{A}_{k}$ are undetermined functions of the
worldsheet time.  
The above expression is also subject to the Virasoro
constraint which implies that, 
\bea 
\lim_{\sigma\rightarrow\sigma_{k}} 
\left[\frac{1}{2}{\rm tr}_{2} [j^{2}_{\pm}(\sigma ,\tau)]\right] & = & 
\lim_{\sigma\rightarrow\sigma_{k}} 
\left[\frac{1}{2}{\rm tr}_{2} [j^{2}_{\tau}(\sigma,\tau)]\right] \,\,=\,\,
\frac{J^{2}}{\sqrt{\lambda}}  \label{vir2} \eea 
where we have used the fact that the space-like component of the current
vanishes at the spike. The above constraint can only be obeyed in
(\ref{comp}) if,  
\bea \eta_{AB} L^{A}_{k} L^{B}_{k} & = & 0 \label{cas2}
\eea 
for each value of $k$. Evaluating the total charge by integrating over
the string, the highest-weight condition becomes, 
\bea  \sum_{k=1}^{K}\,\, \vec{L}_{k}  &  = & \left(\begin{array}{c} 
S \\ 0 \\ 0 \end{array}\right) \label{hw2} \eea
As a check, for $K=2$ we can solve the conditions (\ref{cas2},\ref{hw2}) and 
recover (\ref{l12}) as the general solution. In the general case there
are $2K-2$ remaining free parameters (including $S$), 
as expected from the finite gap construction. 
\paragraph{}
We will now treat the unknown quantities ${L}^{A}_{k}$ as dynamical
variables. We choose a cyclic ordering for the $K$ spikes; 
$0<\sigma_{1}<\sigma_{2}<\ldots<\sigma_{K}<2\pi$ and introduce $K$
arbitrary points on the string $\mu_{k}\in (0,2\pi)$ with, 
$\mu_{k}<\sigma_{k}<\mu_{k+1}$ for $j=1,\ldots,K$ with the 
convention that $\mu_{K+1}=\mu_{1}$. We can then write, 
\bea
L^{A}_{k} & = & \lim_{S \rightarrow \infty} \,\,\left[ 
\frac{\sqrt{\lambda}}{8\pi}\,
\int_{\mu_{k}}^{\mu_{k+1}}\,d\sigma\, j^{A}_{\tau}(\sigma, \tau)\right]
\label{ldef} \eea
\paragraph{}
In the Hamiltonian formalism for the Principal Chiral Model, 
the $\tau$-component of the Noether current has Poisson brackets,  
\bea 
\{j^{A}_{\tau}(\sigma,\tau),\,j^{B}_{\tau}(\sigma',\tau)\} & = & 
-\frac{4\pi}{\sqrt{\lambda}}\,\,2\varepsilon^{ABC}
j_{\tau\, C}(\sigma.\tau)\, \delta(\sigma-\sigma')
\label{pbpcm} \eea 
Substituting (\ref{ldef}) for $j^{A}_{\tau}$ in  (\ref{pbpcm})
we obtain the brackets, 
\bea 
\{L^{A}_{j}, L^{B}_{k}\} & = & -\,\varepsilon^{ABC} \,\delta_{jk}
L_{C\,k} \label{pb3} \eea 
for the variables $L_{k}^{A}$. These steps certainly involve dynamical
assumptions and it remains to be shown that the $L^{A}_{k}$ are not
subject to additional constraints. Another question is whether
we should also include dynamics for the locations, $\sigma_{j}$, of the
spikes. These issues could be addressed by reconstructing actual
string solutions as in \cite{DV1,DV2} and then taking the
large-$S$ limit. For the moment we will rely on the consistent outcome
of this analysis to provide some retrospective 
justification for the assumptions made. 
\paragraph{}
Now we are ready to consider the scaling limit of the monodromy
matrix,\bea 
\Omega\left[x;\tau \right] & = &
\,\,\mathcal{P}\,\,\exp\left[\frac{1}{2}
\int_{0}^{2\pi}\,d\sigma \,\, \left( \frac{j_{+}}{x-1} \,\,+\,\,
\frac{j_{-}}{x+1}\right )\right] 
\nn \eea  
As above 
we have $j^{A}_{\pm}\sim S$ and we also scale the spectral parameter as
$x\sim S$ as $S \rightarrow \infty$. The monodromy matrix
becomes,
\bea 
\Omega\left[x;\tau \right] & \simeq &
\,\,\mathcal{P}\,\,\exp\left[\frac{1}{{x}}
\int_{0}^{2\pi}\,d\sigma \,\,j_{\tau} \,\,\right] \nn \eea 
Using the limit form (\ref{comp}) for ${j}_{\tau}(\sigma,\tau)$
we obtain, 
\bea 
\Omega\left[x;\tau \right] & \simeq & \prod_{k=1}^{K}\,\,\exp\left[ 
\frac{4\pi}{\sqrt{\lambda}}\,\,
\frac{1}{{x}}\, \eta_{AB}L^{A}_{k}s^{B} \,\right]
  \nn \\  & = &
  \frac{1}{{u}^{K}}\prod_{k=1}^{K}\,\, \mathbb{L}_{k}(u) \label{eeq} \eea 
where we set $u=\sqrt{\lambda}x/4\pi$ and identify, 
\bea  
\mathbb{L}_{k}(u) & = & 
\left[u
\mathbb{I}\,\,+\,\,  \eta_{AB}{L}^{A}_{k}s^{B}\right] \nn \\ 
& = & \left(\begin{array}{cc} u+iL^{0}_{k} & L^{1}_{k}+iL^{2}_{k} \\ 
L^{1}_{k}-iL^{2}_{k} & u-iL^{0}_{k} \end{array}\right) \nn  \eea
where we have used the explicit choice form of the generators given
above. Notice that the last equality in (\ref{eeq}) is exact because
the Taylor expansion of the exponential truncates after two terms by
virtue of the relation (\ref{cas2}).  Finally, the above expression
for $\Omega$ coincides (up to an irrelevant overall factor) 
with the monodromy (\ref{monod3}) of the classical 
$SL(2,\mathbb{R})$ spin chain if we identify $\mathcal{L}^{0}_{k}=
L^{0}_{k}$ and $i\mathcal{L}^{\pm}_{k}=L^{1}_{k}\pm iL^{2}_{k}$ 
as the classical spin at the $k$'th site. With this identification we
also reproduce the Poisson brackets (\ref{pb}), 
quadratic Casimir condition (\ref{Cas}) and highest-weight condition 
(\ref{hw}) of the spin chain. 
\paragraph{}
The above analysis indicates that the motion of the spikes 
is governed by the same finite-dimensional complex
integrable as the gauge theory spins. In particular the evolution of
the spikes in global $AdS$ time should be generated by the Hamiltonian 
$H_{K}=\log q_{K}$. It is not quite clear if the relevant trajectories are
literally the same as the depends also on reality conditions for the
initial data. It would be interesting to investigate this further and 
construct some explicit trajectories of the spikes using the methods 
of \cite{KK}.  

\section{Conclusion}
\paragraph{}
In this paper we have argued that the dynamics of the $K$ gap solution
of classical string theory on $AdS_{3}\times S^{1}$ is effectively
described by a classical spin chain of length $K$ in the limit of
large angular momentum, $S\rightarrow \infty$. Thus the continuous
string effectively gives rise a finite-dimensional lattice system 
in the large-$S$ limit. This is the opposite of the usual situation
where a continuous system arises as the thermodynamic or
continuum limit of a discrete one. 
\paragraph{}
Building on the ideas of \cite{Kruc}, we have argued that this new
phenomenon can be understood in terms of the
localisation of the worldsheet fields at $K$ special points or
spikes. Another point of view is provided by the degeneration of the
spectral curve shown in Figure 4. The moduli of the degenerate curve
$\tilde{\Sigma}_{1}$ correspond to the the $K$ lowest modes of the
string\footnote{As mentioned above we are exciting only modes of the
  string which carry positive angular momentum $S$}. The remaining
modes of the string correspond to the double points mentioned in
Section 3 where the quasi-momentum $p(x)$ attains a value $n\pi$ for
some $n\in \mathbb{Z}$. On the initial curve $\Sigma$ these double
points accumulate at the four singular points $\pm 1^{\pm}$. 
In the limit $S\rightarrow \infty$, the singular points and all the
double points end up on the genus zero curve
$\tilde{\Sigma}_{2}$. This has a simple interpretation: the lowest $K$
modes of the string effectively decouple from the infinite tower of
higher modes as $S\rightarrow\infty$ 
and become an isolated finite dimensional system.           
\paragraph{}
Another mysterious aspect of the results presented above is the
 precise matching between one-loop gauge theory and semiclassical 
string theory up
 to a single universal function of the 't Hooft coupling. The
 decoupling described in the previous paragraph throws some light on
 this. Consider the one-loop correction to the semiclassical 
large-$S$ spectrum in the string $\sigma$-model. 
This is calculated by summing the
 small fluctuation frequencies for all the worlsheet fields (including
 fluctuations of all the $AdS_{5}\times S^{5}$ worldsheet
 fields). These frequencies are in turn determined by evaluating a
 particular abelian integral $q(x)$ (the quasi-energy) at each of the
 double points mentioned above \cite{Vicedo}. Because of the factorisation of
 $\Sigma$ into two disjoint 
components $\tilde{\Sigma}_{1}$ and $\tilde{\Sigma}_{2}$, it is easy
 to see that the frequencies which only depend on data determined by  
$\tilde{\Sigma}_{2}$ are independent of the moduli of
 $\tilde{\Sigma}_{1}$. Similar considerations
 should also apply to the fluctuations of the string outside
 $AdS_{3}\times S^{1}$.  It follows that the one-loop correction will be 
the same for all states in the spectrum (\ref{stringspec}). The
 agreement we have found suggests that this argument might
 extend to all $\sigma$-model loops.       
\paragraph{}
It would also be interesting to understand these results in more
detail from the point of view of the planar $\mathcal{N}=4$ theory. 
Spikes near the boundary are dual to localised excitations on $S^{3}$
with the same quantum numbers as an elementary gluon (or other adjoint
field). It would be interesting to investigate possible connections
with the gluon scattering amplitudes discussed in \cite{AlMal}. 
Finally we note that one-loop, large-spin operator spectrum 
(\ref{spectrum1}) is essentially 
universal to all planar four-dimensional gauge theories. 
This suggests that the limit 
of semiclassical string theory studied in this paper may have
applications to large-$N$ QCD.   
\paragraph{}
The author would like to thank Harry Braden, Niklas Beisert, 
Voldya Kazakov, Keisuke Okamura, Matthias Staudacher, Arkady Tseytlin
and Benoit Vicedo for useful discussions. He is also grateful for the
hospitality of the GGI, Florence where much of this work was
completed. 

\section*{Appendix}
\subsection*{Matching condition}
\paragraph{}
On the original spectral curve 
\bea 
\Sigma\,\,:\qquad{} y^{2} & = & \left(x-b\right)  
\left(x+b\right) \prod_{i=1}^{K-1} 
\left(x-a^{(i)}_{+}\right)\left(x-a^{(i)}_{-}\right)
\nn \eea
the extra A-cycle condition can be written as, 
\bea \oint_{\mathcal{A}^{+}_{K/2}}\,dp \,\,=\,\, 2I_{1} \,+ \,2I_{2} & =
& 0 \label{conds} \eea 
where, 
\bea I_{1}\,=\, \int_{b}^{a_{+}^{(1)}}\,\, dp_{1} & \qquad{} & 
I_{2}\,=\, \int_{b}^{a_{+}^{(1)}}\,\, dp_{2} \nn \eea 
with the explict expressions for $dp_{1}$ and $dp_{2}$ given in 
Eqn (\ref{ansatz}).   
\paragraph{}
The first integral can be treated using the change of variables
$x=\rho \tilde{x}$ which gives,  
\bea y^{2} & = & \rho^{2K}
\left(\tilde{x}^{2}-\frac{b^{2}}{\rho^{2}}\right) \tilde{y}_{1}^{2} 
\nn \eea
where $\tilde{y}_{1}$ is the hyperelliptic coordinate on the curve
$\tilde{\Sigma}_{1}$, 
\bea 
\tilde{\Sigma}_{1}\,\,:\qquad{} \tilde{y}^{2}_{1} & = & \prod_{i=1}^{K-1} 
\left(\tilde{x}-\tilde{a}^{(i)}_{+}\right)\left(\tilde{x}-
\tilde{a}^{(i)}_{-}\right)
\nn \eea
 Then we have, 
\bea I_{1} & = & -\,\int_{\epsilon}^{\tilde{a}^{(1)}_{+}} \,\,
\frac{\left(\sum_{\ell=0}^{K-2}\,\tilde{C}_{\ell}\,\tilde{x}^{\ell}\right)}
{\tilde{y}_{1}}\,\,\frac{d\tilde{x}}{\sqrt{\tilde{x}^{2}-\epsilon^{2}}}
\nn \eea 
where $\epsilon=b/\rho$. We need to find the leading behaviour of this
integral as $\epsilon\rightarrow 0$. For this purpose it is convenient
to write, 
\bea 
I_{1}\,\,=\,\,\frac{1}{\epsilon}\frac{\partial}{\partial\epsilon}\,\, 
\hat{I}(\epsilon) & {\rm where} &  \hat{I}(\epsilon)\,\,=\,\,
\int_{\epsilon}^{\tilde{a}^{(1)}_{+}}
\,\,\sqrt{\tilde{x}^{2}-\epsilon^{2}}\,\,\frac{\left(\sum_{\ell=0}^{K-2}
\,\tilde{C}_{\ell}\,\tilde{x}^{\ell}\right)}
{\tilde{y}_{1}}\,d\tilde{x} \nn \eea 
and expand the suqare root in the integrand in powers of $\epsilon^{2}$. 
\bea
\hat{I}(\epsilon)\,\,=\,\, \sum_{k=0}^{\infty} \,\,\epsilon^{2k} \,\, 
\hat{I}_{k}(\epsilon) \nn \eea 
with, 
\bea 
\hat{I}_{k}(\epsilon) & = & (-1)^{k}\, \left(\begin{array}{c} \frac{1}{2} \\ k 
\end{array}\right) \,\, \int_{\epsilon}^{\tilde{a}^{(1)}_{+}}
\,\,
\frac{\tilde{x}^{1-2k}\left(\sum_{\ell=0}^{K-2}
\,\tilde{C}_{\ell}\,\tilde{x}^{\ell}\right)}
{\tilde{y}_{1}}\,d\tilde{x} \nn \eea 
Each term $\hat{I}_{k}(\epsilon)$, with $k\neq 1$, is analytic in
$\epsilon$ the leading contribution as $\epsilon\rightarrow 0$ is
proportional to $\tilde{C}_{0}/\tilde{Q}=K$ ($\tilde{Q}$ is defined
in Eqn (\ref{Qtilde}) above). As a result each of these terms only gives
rise to a moduli independent constant in the $\epsilon\rightarrow 0$
limit. The leading moduli-dependence comes from the remaining term 
$\hat{I}_{1}(\epsilon)$ which is non-analytic at $\epsilon=0$. The
resulting contribution to $I_{1}$ is,  
\bea 
I_{1}\,\,\simeq\,\,\frac{1}{\epsilon}\frac{\partial}{\partial\epsilon}
\,\, \epsilon^{2}\,\hat{I}_{1}(\epsilon) & \simeq & -\,  
\int_{\epsilon}^{\tilde{a}^{(1)}_{+}}
\,\,
\frac{\left(\sum_{\ell=0}^{K-2}
\,\tilde{C}_{\ell}\,\tilde{x}^{\ell-1}\right)}
{\tilde{y}_{1}}\,d\tilde{x} \nn \eea  
The remaining integral can be then expressed as a contour integral on
the curve $\tilde{\Sigma}_{1}$, 
 \bea 
I_{1} & \simeq & \frac{1}{2}
\int_{\epsilon^{-}}^{\epsilon^{+}}\, d\tilde{p}_{1} \,\,=\,\, 
\tilde{p}_{1}(\tilde{x}=\epsilon)
\eea 
with $\epsilon=b/\rho$, where $\epsilon^{\pm}$ are the points above
$\tilde{x}$ on $\tilde{\Sigma}_{1}$. 
Using the explicit formula 
$\tilde{p}_{1}(\tilde{x})=\cos^{-1}(f/2)$ with $f$ given as given in
Eqn (\ref{f}), 
\bea
I_{1}  & \simeq & \frac{1}{i} \log 
\left(\frac{\hat{q}_{K}\rho^{K}}{b^{K}}\right)\,\, +\,\, \ldots \nn
\eea
where the dots denote subleading terms.  
\paragraph{}
The second integral $I_{2}$, in the period condition (\ref{conds}) 
has limiting behavior, 
\bea 
I_{2} & = & \int_{b}^{\infty} \, d\tilde{p}_{2} \nn  
 \\ & = & -\,\frac{i\pi J}{\sqrt{\lambda}}\int_{b}^{\infty}\left[ 
\sqrt{b^{2}-1}\left(\frac{1}{(x-1)^{2}}\,+\,\frac{1}{(x+1)^{2}}\right)
\,\,-\,\, 
\frac{1}{\sqrt{b^{2}-1}}\left(\frac{1}{(x-1)}\,-\,\frac{1}{(x+1)}\right)
\,\right]\,\,\frac{dx}{\tilde{y}_{2}}\nn  \eea
with, 
\bea 
\tilde{y}^{2}_{2} & = &
x^{2}\,-\,b^{2} 
\nn \eea
Anticipating the fact that $b\rightarrow 1$ as $\rho\rightarrow 0$, 
the leading piece is, 
\bea 
I_{2} & \simeq & 
\frac{2\pi J
  i}{\sqrt{\lambda}}\,\,\frac{1}{\sqrt{b^{2}-1}}\,\,+\,\,\ldots  \nn
\eea 
where the dots denote subleading terms.

\end{document}